# The Cosmic Baryon Cycle in IllustrisTNG: Flows of Mass, Energy, and Metals

Yossi Oren[1], Viraj Pandya[2], Rachel S. Somerville[3], Shy Genel[2,3], Osase Omoruyi[4], and Amiel Sternberg[1,3,5]
[1] School of Physics and Astronomy, Tel Aviv University, Ramat Aviv 69978, Israel; orenyossi01@gmail.com
[2] Columbia Astrophysics Laboratory, Columbia University, 550 West 120th Street, New York, NY 10027, USA
[3] Center for Computational Astrophysics, Flatiron Institute, 162 5th Avenue, New York, NY, 10010, USA
[4] Center for Astrophysics, Harvard & Smithsonian, 60 Garden Street, Cambridge, MA 02138, USA
[5] Max-Planck-Institut für extraterrestrische Physik (MPE), Giessenbachstraße, 85748 Garching, Germany
*Received 2025 October 6; revised 2026 January 20; accepted 2026 February 2; published 2026 March 9*

## Abstract

We measure and analyze the inflows and outflows of mass, energy, and metals through the interstellar medium (ISM) and circumgalactic medium (CGM) of galaxies in the TNG100 simulation. We identify the dominant feedback mechanism in bins of halo virial mass and redshift by computing the integrated energy input from supernovae (SNe) and the "kinetic" and "thermal" modes of active galactic nucleus (AGN) feedback. We measure all quantities in a shell at the virial radius ("halo scale") and one chosen to be approximately at the interface of the CGM and the ISM ("ISM scale"). At $z \gtrsim 2$, we find that galaxies have strong net positive inflows on halo scales, and weaker but still net positive inflows on ISM scales. At later times, partially due to the onset of kinetic AGN feedback in massive halos, inflows and outflows nearly balance one another, leading to the familiar effects of the slowdown of galaxy growth and the onset of quenching. Halos dominated by SN feedback show only weak evidence of preventative feedback on halo scales, and we see excess ISM-scale accretion indicative of rapid gas recycling. Wind mass loadings decrease with increasing halo mass, and with increasing redshift, while energy loadings are nearly independent of both mass and redshift. The detailed measurements of these mass, metal, and energy inflow and outflow rates on galaxy and halo scales can be used to guide empirical and semianalytic models, and provide deeper insight into how galaxy growth and quenching are regulated in the IllustrisTNG simulations.

## 1. Introduction

Our current understanding of galaxy evolution is based on the Λ cold dark matter (ΛCDM) model for hierarchical structure formation (S. D. M. White & M. J. Rees 1978; G. R. Blumenthal et al. 1984; M. Davis et al. 1985). It posits that gravitational instability drives the collapse of DM-dominated overdensities, along with the associated baryons, into bound structures. These dark matter (DM) halos formed the sites where gas was able to cool and form the first stars. Halos grow continuously over time via the accretion of gas and DM as well as via mergers, as larger and larger scales become gravitationally unstable and collapse.

From the first theoretical attempts to account for the properties of galaxy populations within the ΛCDM framework, it has been clear that some kind of "feedback" is required across the spectrum of halo masses (S. D. M. White & M. J. Rees 1978; A. Dekel & J. Silk 1986; S. D. M. White & C. S. Frenk 1991). Not all of the accreted gas that is able to cool rapidly can be allowed to form stars efficiently, or both low-mass and high-mass galaxies would be overproduced relative to observations. Another way to cast this is in terms of the "baryon conversion efficiency," or the fraction of a halo's "baryon budget" that is in the form of stars ($m_*/(f_B M_{vir})$, where $m_*$ is the mass in stars, $f_B$ is the universal baryon fraction, and $M_{vir}$ is the mass of the halo). Observational constraints from abundance matching, gravitational lensing, and galaxy clustering show that this baryon conversion efficiency is quite low over most of cosmic history ($z \sim 0\%$–10%), reaching a maximum value of ∼20% at its peak and dropping to only a few percent in the lowest and highest halo masses that host observable galaxies (B. P. Moster et al. 2010; P. S. Behroozi et al. 2013; R. H. Wechsler & J. L. Tinker 2018; P. Behroozi et al. 2019). This implies that a large fraction of the gas that would be expected to cool and accrete into the interstellar medium (ISM) and form stars in the absence of feedback does not do so, either because it is prevented from accreting and/or cooling as rapidly as expected (preventative feedback) or because mass is efficiently ejected from the ISM and/or circumgalactic medium (CGM; ejective feedback).

There is also direct evidence that stars and supermassive black holes (SMBHs) deposit copious amounts of energy and momentum into their surroundings (T. M. Heckman & P. N. Best 2023, and references therein). Decades of observations of "galactic winds" show large-scale outflows (e.g., D. E. Osterbrock 1960; E. M. Burbidge et al. 1964; S. Veilleux et al. 2005; D. S. N. Rupke 2018). Galactic winds can extend well beyond the ISM and into the CGM (e.g., M. Yoshida et al. 2016; E. Falgarone et al. 2017; Y. Guo et al. 2023), and in some cases may even be powerful enough to overcome the binding energy of the halo and reach the intergalactic medium (IGM; D. S. N. Rupke et al. 2019).

Two primary mechanisms drive these large-scale outflows: supernova (SN) explosions and active galactic nuclei (AGN). Massive stars and SNe deposit energy and momentum into the ISM, driving an overpressured bubble of hot gas that can entrain cold material and vent mass, energy, and metals out of the galaxy (e.g., D. K. Strickland & T. M. Heckman 2009; D. S. N. Rupke 2018). AGN can drive outflows via two mechanisms. Rapid accretion onto the black hole (BH) is thought to lead to efficient production of radiation (associated with classical quasars and AGN), which can drive winds via radiation pressure, magnetic pressure, and line pressure (sometimes called "quasar mode" or "bright mode"). BHs accreting at lower rates can drive powerful relativistic jets, associated with radio galaxies (sometimes called "radio mode" or "jet mode;" A. C. Fabian 2012; T. M. Heckman & P. N. Best 2014). Observationally, it is often difficult to disentangle whether a wind is driven by SNe or an AGN, and many winds may have contributions from both types of sources (S. Veilleux et al. 2005). However, it is likely that winds in lower-mass galaxies, which tend to be star forming and may have lower-mass BHs, are primarily driven by SNe, and observed outflows in more massive galaxies (particularly "quenched" galaxies with little ongoing star formation) are driven by AGN. These wind processes occur across a wide range of spatial and temporal timescales. The energy and momentum deposition from SNe and BHs occur on physical scales of less than a parsec, while galactic winds and BH jets can extend out to hundreds of kiloparsecs. Stars that evolve over millions of years release SN energy in seconds, and BH accretion rates can vary by orders of magnitude on timescales of months to years.

The processes of intergalactic gas inflow, ejection via feedback, and reaccretion onto the galaxy are collectively termed "the baryon cycle," and it is viewed as crucial to our understanding of galaxy evolution (Astro2020 Decadal Report).[6] The vast range of spatial and temporal scales, and the importance of a wide range of complex, nonlinearly interconnected physical processes including gravity, hydrodynamics, thermodynamics, magnetic fields, and radiation, make modeling the baryon cycle in a cosmological context extremely challenging (R. S. Somerville & R. Davé 2015). Numerical cosmological simulations approach this task by explicitly solving the relevant equations numerically on a (sometimes adaptive or moving) mesh or using Lagrangian, particle-based methods such as smoothed particle hydrodynamics (M. Vogelsberger et al. 2020). Many of the key processes that shape galaxy evolution, including star formation, stellar and SN feedback, and BH seeding, accretion, and feedback occur on spatial scales below the resolution that is achievable in large volume cosmological simulations (typically hundreds of parsecs to $\sim$1 kpc). As a result, these processes are modeled using "subgrid" recipes, which historically have been implemented in a somewhat heuristic, phenomenological manner (see the discussion in R. S. Somerville & R. Davé 2015; T. Naab & J. P. Ostriker 2017). These recipes contain free parameters that reflect our ignorance about the details of how the smaller-scale phenomenon manifest on the coarse-grained scale, which are commonly tuned to reproduce certain quasi-observable global galaxy properties at the population level, such as the galaxy stellar mass function. Different subgrid implementations can give rise to very different predictions for quantities that have not been explicitly calibrated, particularly properties of difficult-to-observe diffuse gas in the CGM and IGM (M. T. Tillman et al. 2023; R. J. Wright et al. 2024). Furthermore, different subgrid implementations can arrive at similar $z \sim 0$ galaxy stellar properties via very different routes, exhibiting rather dramatically different baryon cycles (V. Pandya et al. 2020; R. J. Wright et al. 2024).

In spite of its fundamental importance, many basic questions about feedback and the baryon cycle in these large cosmological simulations remain open. These include: Is feedback primarily ejective or preventative? How far does ejected material travel, and how many times does it get recycled back into the ISM? What is the specific energy (or ratio of energy loading to mass loading) in winds? Which feedback processes (SN-driven winds, or radiatively efficient or inefficient AGN) are important at different times, scales, and halo masses? How do metal flows trace baryon flows? What is the relationship between wind launching parameters and emergent properties of outflows on larger scales?

Although numerical cosmological simulations are a powerful tool, they are computationally expensive, making it difficult to simulate large volumes at adequate resolution, and to explore the effects of varying the subgrid parameters or implementations. A complementary approach to modeling the baryon cycle is a technique that uses semianalytic models (SAMs), in which galaxies are represented by a collection of reservoirs such as the stellar bulge and disk, ISM, CGM, etc., set within "merger trees" that represent the cosmological growth of halos through accretion and merging (e.g., G. Kauffmann et al. 1993; R. S. Somerville & J. R. Primack 1999; S. Cole et al. 2000; A. J. Benson 2010). Simplified models of physical processes such as gas accretion, heating and cooling, star formation, chemical enrichment, and stellar and BH feedback are used to track the flows of material between these different reservoirs by solving a system of nonlinearly coupled ordinary differential equations (ODEs). An important recent development is the addition of ODEs describing the energy flows between reservoirs, in addition to mass and metals as in traditional SAMs. C. Carr et al. (2023) and V. Pandya et al. (2023) showed that energy flows can play a critical role in regulating cooling and accretion from the CGM to the ISM, thereby effectively regulating star formation without requiring large amounts of mass to be ejected from the ISM (see also G. M. Voit et al. 2024a, 2024b).

Traditionally, the scaling relations adopted in SAMs to describe processes such as star formation or stellar-driven winds were largely empirical or phenomenological, and were parameterized and tuned to match observational constraints. However, more recently, there have been attempts to implement scalings extracted from hydrodynamic simulations within SAMs (V. Pandya et al. 2023) or to use them to motivate the scalings used in closely related "gas regulator" models (P. D. Mitchell & J. Schaye 2022). It is an interesting open question whether in this way the SAM framework can effectively be used as a physically interpretable "emulator" for more complex and more expensive numerical simulations.

The goal of this work is to measure the detailed dependence of the inflow and outflow rates of mass, metals, and energy, on both CGM and ISM scales, on halo mass and redshift in the TNG100 simulation from the IllustrisTNG suite of numerical cosmological simulations (A. Pillepich et al. 2018a; D. Nelson et al. 2019a). In a companion paper, we use these results to inform and benchmark a new SAM that attempts to reproduce the predictions of IllustrisTNG not only for the "state variables" (galaxy properties such as stellar mass, gas mass, metallicity, etc.) but also flow rates between different

[6] https://nap.nationalacademies.org/catalog/26141/pathways-to-discovery-in-astronomy-and-astrophysics-for-the-2020s

reservoirs, over a broad range in halo mass and redshift (O. Omoruyi et al. 2026, in preparation).

Previous works have described aspects of the baryon cycle in and around simulated galaxies. For example, D. Nelson et al. (2015) tested the impact of feedback on cosmological gas accretion in the original Illustris simulation; D. Nelson et al. (2019b) investigated outflows in TNG50 galaxies; E. Zinger et al. (2020) studied how the onset of AGN feedback in TNG100, particularly the low-accretion, "jet-mode" feedback, affects galaxies and their CGM; V. Pandya et al. (2021) performed an analysis of mass, momentum, energy, and metal outflows in the FIRE-2 simulation suite; P. D. Mitchell et al. (2020a, 2020b) described the baryon cycle in the EAGLE simulation; R. J. Wright et al. (2024) tracked the mass flow rates in and out of galaxies in several cosmological simulations, including IllustrisTNG; and J. Morgan et al. (2025) analyzed the flow of mass and metals in and out of isolated TNG100 galaxies. In this paper, we build on these works by extracting both inflow and outflow rates, on scales chosen to match the traditional reservoirs used in SAMs (ISM and CGM), and tabulate these results over a more finely binned grid in halo mass and redshift than previous works. Another novel aspect of our analysis is that in addition to mass and metallicity, we also measure the energy flows on both ISM and halo scales, which has not been presented previously for the IllustrisTNG simulations. We present our results separately for galaxies that are dominated by SN feedback and those that are AGN dominated, which has not generally been done in previous studies. In addition to being useful for developing next-generation SAMs, our results provide numerous insights into the physics of how both stellar and AGN feedback shape the properties of galaxies and their CGM in the IllustrisTNG universe.

The structure of our paper is as follows: In Section 2 we describe the IllustrisTNG simulation suite and the galaxy sample we use for our analysis; in Section 3 we describe in-depth the way feedback is implemented in The Next Generation (TNG) simulation, and determine the dominant feedback mechanism in TNG halos as a function of mass and redshift; in Section 4 we describe how we measure the flow of mass, energy, and metals, alongside defining loading factors; in Sections 5, 6, and 7 we analyze the flow of mass, energy, and metals in and out of TNG galaxies and halos; in Section 8 we discuss the interpretation of our results; and finally in Section 9 we present our summary and conclusions.

## 2. The IllustrisTNG Simulations

The IllustrisTNG simulation is an advanced cosmological hydrodynamical simulation designed to study galaxy formation, large-scale structures, and the interaction between dark and baryonic matter (F. Marinacci et al. 2018; J. P. Naiman et al. 2018; D. Nelson et al. 2018; A. Pillepich et al. 2018b; V. Springel et al. 2018). It uses the AREPO code (V. Springel 2010), which adopts a moving mesh finite-volume method, to solve coupled equations of magnetohydrodynamics, and a tree-based approach for gravity (R. Pakmor et al. 2011; R. Pakmor & V. Springel 2013).

To account for physical phenomena unresolved at the simulation scale, IllustrisTNG includes several subgrid models. Star formation is modeled with a density-dependent prescription, where dense, cold gas can form stars; a chemical enrichment model tracks the evolution of elements in stars and their distribution in the surrounding gas; stellar feedback in the form of SNe injects energy into the surrounding gas, driving galactic winds; radiative mechanisms allow gas heating in the presence of the universal background and local AGN radiation, alongside metal-line cooling; and SMBHs are formed, grow, and deposit energy into their surroundings in the form of thermal and kinetic AGN feedback (R. Weinberger et al. 2017; A. Pillepich et al. 2018a). In Section 3 we dive deeper into the implementation of feedback via these subgrid processes.

The IllustrisTNG simulation suite includes three box sizes: 51.7, 110.7, and 302.6 Mpc on a side, each with a different resolution marked "1" (highest) to "3" (lowest). In this work we use TNG100-1, which is the highest resolution available for the 110.7 Mpc box size. This run includes $1820^3$ DM particles with a mass of $7.5 \times 10^6\,M_\odot$, and gas particles with a mass of $1.4 \times 10^6\,M_\odot$, which form stars in a similar mass resolution. As DM halos and galaxies form in the simulation, their identification is carried out using two algorithms: the Friends-of-Friends (FOF) algorithm is used to identify DM halos, while the SUBFIND algorithm (V. Springel et al. 2001) identifies substructures in the formed halos, i.e., galaxies. To track the evolution of galaxies over time, the SUBLINK algorithm (V. Rodriguez-Gomez et al. 2015) is used to construct merger trees at the subhalo or galaxy level.

The cosmological parameters used in IllustrisTNG are based on the Planck Collaboration et al. (2016) results: a Hubble constant $H_0 = 67.74\ \mathrm{km\,s^{-1}\,Mpc^{-1}}$, matter and baryon density parameters $\Omega_{m,0} = 0.3089$ and $\Omega_{\mathrm{B},0} = 0.0486$, respectively, and a cosmological constant $\Omega_{\Lambda,0} = 0.6911$. The simulation evolves the universe from high redshift ($z = 127$) to the present ($z = 0$), spanning 100 snapshots. At $z = 0$, there are over $6 \times 10^6$ identified DM halos containing a total of over $4 \times 10^6$ galaxies.

### 2.1. Halo Sample

To perform a robust analysis of galactic inflows and outflows and understand their influence over cosmic time, our analysis must span a wide range of redshifts and virial masses. We therefore analyze halos taken from 10 redshifts, each a different snapshot of the simulation: 0, 0.1, 0.2, 0.5, 1, 2, 4, 6, 8, and 10.

At each redshift, we randomly select up to 100 halos in bins of 0.3 dex in virial mass, starting from $M_{\mathrm{vir}} = 10^{10}\,M_\odot$ and reaching up to the most massive halo in each snapshot. In total, across all redshifts we analyze 9522 halos. In Figure 1 we present the number of halos per redshift and virial mass bin. Note that because we pick halos randomly at each redshift, we do not track the direct evolution of individual halos along a common merger tree, unlike V. Pandya et al. (2020, 2021, 2023).

For each halo we collect all gas, DM, BH, and star particles within 1.5 virial radii from the point of the minimum gravitational potential energy inside the halo. While we use the FOF and SUBFIND algorithms to identify halos and their central galaxies, we do not collect particles based on their association with each FOF halo, but pick particles based on their location.[7] We have found that as we reach higher fractions of the virial radius, a significant portion of the particles that can be found around the halo are not associated with its FOF group. To analyze the feedback history in each halo, we additionally track the main progenitor branch of its central galaxy using the SUBLINK merger trees.

[7] A similar particle selection process was used by J. Morgan et al. (2025).

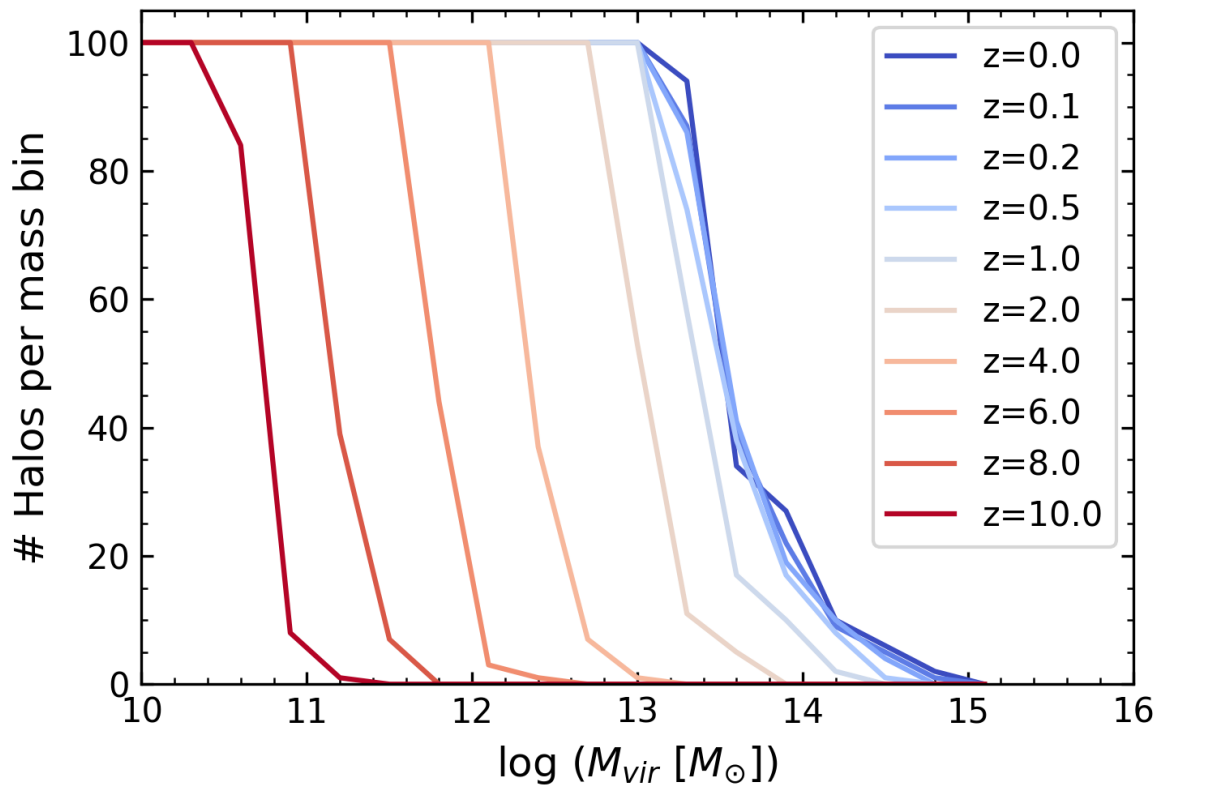

**Figure 1.** The distribution of TNG100 halos analyzed in this work as a function of halo virial mass, colored by redshift. In total, 9522 TNG100 halos are analyzed across 10 redshifts and in virial mass bins of 0.3 dex, starting from $10^{10}$ $M_\odot$ and up to the most massive halo in each snapshot.

When discussing the virial mass we refer to $M_{\Delta,c}$, using the G. L. Bryan & M. L. Norman (1998) definition of the overdensity relative to the critical density of the universe. Appropriately, we define the virial radius as $R_{\Delta,c}$. We define the ISM radius as $0.1R_{\rm vir}$, in accordance with previous works (e.g., A. L. Muratov et al. 2015; D. Anglés-Alcázar et al. 2017a; P. D. Mitchell et al. 2020a; V. Pandya et al. 2021; Z. Hafen et al. 2022). As for galactic properties, we take the radius of the galaxy to be twice its stellar half-mass radius, inside of which we estimate properties such as stellar mass or star formation rate (SFR).[8]

A significant point to consider is whether the simulation results we analyze are well resolved. Unresolved galaxies can be dominated by shot noise that increases with fewer particles. It is common to consider resolved galaxies as those that have more than 100 particles, corresponding to a shot noise of under 10% (e.g., A. Pillepich et al. 2018a; D. Nelson et al. 2019b; P. D. Mitchell et al. 2020b). In TNG100, this limit translates to stellar masses $\sim 10^8\,M_\odot$, equivalent to a virial mass of $\sim 10^{10.8}\,M_\odot$. As the mass, energy, and metal outflow rates we measure are driven by galactic subgrid processes, in our results we use a gray band to highlight all results from halos with $M_{\rm vir} < 10^{11}\,M_\odot$, or our first four mass bins, to indicate their central galaxies may be dominated by shot noise.

An important distinction to make between galaxies in our sample in the context of their SFR is whether they are considered "star forming" or "quenched." In Figure 2 we plot the total instantaneous SFRs inside the central galaxy associated with each halo, for four of the redshifts we analyze in this work. We mark halos whose central galaxy is star forming in blue, and halos whose central galaxy is quenched in red. We determine which is which using the method described in T. Karmakar et al. (2023): for each redshift, we estimate the median specific SFR (sSFR) for galaxies with stellar masses in the ranges $\log(M_*/M_\odot) = 9.25 \pm 0.25$ and $\log(M_*/M_\odot) = 9.75 \pm 0.25$. We draw a line on the $\log(M_*) - \log(\text{sSFR})$ plane between these median sSFRs and the centers of the stellar mass bins to which they correspond, which is then extrapolated to higher and lower masses. Galaxies with an sSFR higher than 25% of the value on this line (according to their stellar mass) are considered star forming. We mark the median SFR per virial mass bin with a black dashed line. As described in previous works such as J. J. Davies et al. (2020) or E. Zinger et al. (2020), the quenching of star formation above the virial masses of halos surrounding $L^*$ galaxies in TNG100 is associated with AGN feedback starving galaxies of their star-forming fuel by heating and ejecting significant amounts of gas from their ISM. That being said, to better understand the effects of feedback on the galaxy and on the halo in which it resides, we need to further discuss how the feedback is implemented.

[8] This radius is included in the public TNG catalogs.

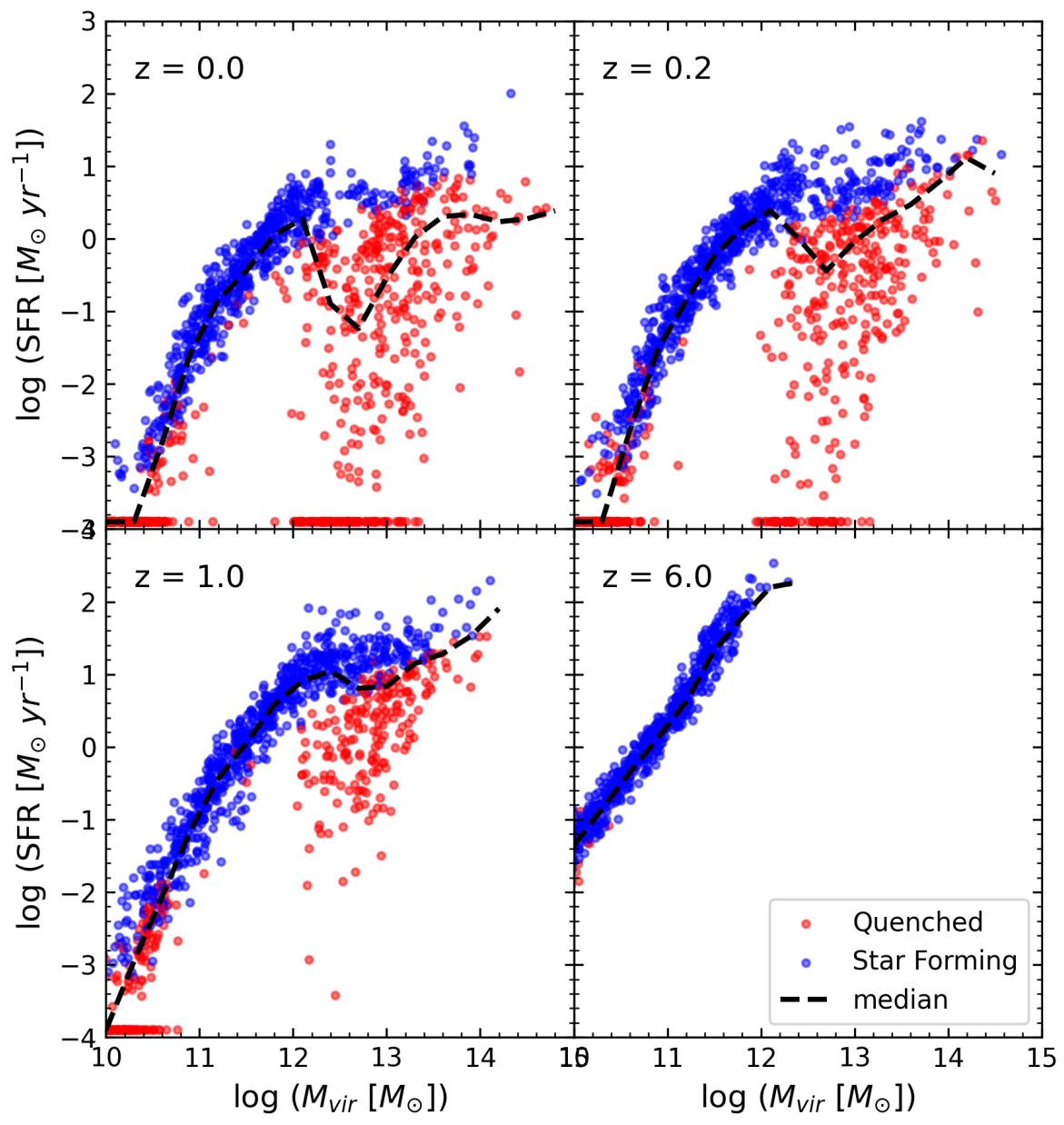

**Figure 2.** The instantaneous SFRs in the central galaxies of all halos from four of the redshifts analyzed in this work (top left: $z = 0$; top right: $z = 0.2$; bottom left: $z = 1$; and bottom right: $z = 6$), as a function of halo virial mass. Halos whose central galaxies are star forming are marked in blue, and halos whose central galaxies are quenched are marked in red (classification into star forming or quenched is based on the method introduced in T. Karmakar et al. 2023). At each redshift, the median is plotted using a dashed black line.

## 3. Feedback in TNG100

In the TNG100 simulation, winds are driven by SNe and AGN, both implemented as subgrid recipes. The implementation of star formation is described in V. Springel & L. Hernquist (2003) and M. Vogelsberger et al. (2013), and the launching of winds due to SNe is described in A. Pillepich et al. (2018a). The seeding of BH particles in growing halos is also described in A. Pillepich et al. (2018a), and the growth and feedback modes from BH particles are described in R. Weinberger et al. (2017). The numerical values shown in the following two sections are all taken from the fiducial, publicly available, TNG100 run.

### 3.1. Star Formation Rate and Supernova Feedback

Star formation in the TNG simulation is implemented stochastically. Any gas particle above a certain density threshold has a probability to form stars in a rate determined by a characteristic star formation timescale. Assuming a G. Chabrier (2003) initial mass function (IMF), a fraction of the formed stellar mass then explodes as SNe, which launch wind particles into the surrounding ISM. These wind particles are launched in random directions with an initial velocity that is given by

$$v_w = \max\left[7.4\,\sigma_{\rm DM}\left(\frac{H_0}{H(z)}\right)^{1/3},\ 350\ {\rm km\ s^{-1}}\right], \quad (1)$$

where $\sigma_{\rm DM}$ is the local DM velocity dispersion. Notice that the wind velocity is capped at a maximum value of $350\,{\rm km\,s^{-1}}$. The rate of wind mass flow per unit SFR is determined by a mass loading factor, given by

$$\eta_{w,\,{\rm TNG}} = \frac{2}{v_w^2} e_w \times 0.9. \quad (2)$$

The multiplication of the mass loading factor by a factor of 0.9 implies that 90% of the energy of the formed wind particle is kinetic, and the remaining 10% is thermal. The parameter $e_w$ is a specific energy that is attributed to a wind particle depending on the metallicity of the star-forming gas cell and on a Type II SN rate per unit mass of stars formed, calculated assuming a G. Chabrier (2003) IMF:

$$e_w = \left[1 + \frac{3}{1 + (Z/0.002)^2}\right] \times 1.06 \times 10^{49}\ {\rm erg}\ M_\odot^{-1}. \quad (3)$$

Thus, the specific energy ranges between $e_w = 4.25 \times 10^{49}\ {\rm erg}\ M_\odot^{-1}$ for metallicities far below the $Z = 0.002$ threshold (equivalent to $\sim$16% of $Z_\odot$) and $e_w = 1.06 \times 10^{49}\ {\rm erg}\ M_\odot^{-1}$ for metallicities far above it. The newly formed wind particle has the following energy rate:

$$\dot{E}_{\rm SNe} = e_w \times {\rm SFR}. \quad (4)$$

The formed wind particles are decoupled from the gas that spawned them, until leaving the ISM and recoupling with the surrounding gas, depositing their momentum, mass, metals, and thermal energy content. The metallicity of a newly formed wind particle is 40% of its parent gas cell, as described in M. Vogelsberger et al. (2013) for the Illustris simulation, and is adopted for IllustrisTNG as well. For a more detailed description of the implementation of SN feedback in IllustrisTNG, see Section 2.3.2 of A. Pillepich et al. (2018a). In our analysis of flow rates, we track stellar wind particles as if they were gas particles.

### 3.2. Black Hole Growth and Active Galactic Nucleus Feedback

In the simulation, a BH seed mass of $1.2 \times 10^6\,M_\odot$ is created and placed in the gravitational potential minimum of an FOF halo that has reached a virial mass threshold of $M_{200,c} = 7.4 \times 10^{10}\,M_\odot$. Each BH particle in TNG is allowed to accrete gas at the Bondi accretion rate, which is capped at the Eddington accretion rate. Unlike stellar feedback, AGN feedback in TNG comes in one of two states, determined by $\dot{M}_{\rm Bondi}/\dot{M}_{\rm Eddington}$, the Bondi accretion rate of the BH particle relative to its Eddington accretion rate. The threshold for delineating between the two feedback states is given by

$$\chi = \min\left[2 \times 10^{-3}\left(\frac{M_{\rm BH}}{10^8\,M_\odot}\right)^2,\ 0.1\right]. \quad (5)$$

As the mass of the BH particle reaches $\sim 7 \times 10^8\,M_\odot$ (the median BH mass in $\sim 10 \times 10^{13}\,M_\odot$ halos), the value of $\chi$ is capped at 0.1.

If $\dot{M}_{\rm Bondi}/\dot{M}_{\rm Eddington} \geqslant \chi$, the BH is assumed to be in a "high-accretion" state, in which pure thermal energy is injected in a the region around the BH particle. The rate of energy injection is

$$\dot{E}_{\rm th,AGN} = 0.02\,\dot{M}_{\rm BH} c^2. \quad (6)$$

If $\dot{M}_{\rm Bondi}/\dot{M}_{\rm Eddington} < \chi$, the BH is in a "low-accretion" state. In this state, the injected energy rate is 10 times higher:

$$\dot{E}_{\rm kin,AGN} = 0.2\,\dot{M}_{\rm BH} c^2. \quad (7)$$

The energy released by BHs in the "low-accretion" mode is accumulated until reaching a threshold of

$$E_{\rm inj,min} = 10\sigma_{\rm DM}^2 m_{\rm enc}, \quad (8)$$

where $m_{\rm enc}$ is the mass enclosed in the feedback region. Past this threshold, a feedback "event" occurs where the energy is injected into the surrounding gas in the form of pure kinetic energy. Per feedback event, gas particles around the BH are given momentum in a random direction. For a more detailed description of the way AGN feedback is implemented in IllustrisTNG, see Section 2 of R. Weinberger et al. (2017). In this work we interchangeably refer to the AGN "high-accretion" state as "thermal feedback," and the "low-accretion" state as "kinetic feedback."

The transition from thermal to kinetic AGN feedback depends on the BH mass such that a newly formed BH particle is in the thermal state, and the likelihood for a BH particle to be in the kinetic state increases with mass. In fact, R. Weinberger et al. (2017) show that even before reaching the mass relevant to the maximum value of $\chi$, most BHs have transitioned from thermal to kinetic feedback. As kinetic feedback drives gas away from the BH more effectively, it also limits the gas available for accretion by the BH particle, further reducing its accretion rate and preventing it from getting back to a thermal feedback mode.

### 3.3. Dominant Feedback Mechanism

In Figure 3 we plot the median instantaneous energy injection rates for all three mechanisms (SNe (thermal + kinetic), thermal AGN feedback, and kinetic AGN feedback) as a function of virial mass and for different redshifts. Scattered points are colored by their redshift, and error bars represent the standard deviation for the mass bin, equivalent to the 16th–84th percentile range (this is true for all error bars presented in this work). While SN feedback occurs for all masses and redshifts, no AGN feedback occurs for $M_{\rm vir} \lesssim 4 \times 10^{10}\,M_\odot$ due to the lower limit on the halo mass for BH seeding in TNG. Due to the way the AGN feedback subgrid recipe is implemented, TNG galaxies generally experience either thermal or kinetic feedback, but not both simultaneously. The relative importance of kinetic and thermal AGN feedback depends on both mass and redshift. We find that the criteria for kinetic AGN feedback are only met at $z \lesssim 2$ and for $M_{\rm vir} \gtrsim 10^{12}\,M_\odot$, while thermal AGN feedback occurs at all redshifts, but only in the mass range of $4 \times 10^{10} \lesssim M_{\rm vir}/M_\odot \lesssim 2 \times 10^{13}$.

We wish to determine which feedback mechanism is more significant in driving outflows at a given time. As a proxy to this question, we compare the energies injected into the ISM by each mechanism. To prevent bias due to peaks in the instantaneous energy injection rates, we estimate for each halo in our analysis the total energy injected by each mechanism over time periods that correspond to three typical travel times from the galaxy to the virial radius. We measure the distribution of outflow velocities at the virial radius of each halo at the redshift of interest, and record the 10th percentile, 50th percentile, and 90th percentile velocities. Using the virial radius at the redshift of interest and under the simplifying assumption that the gas moves at a constant velocity from its launching point, we estimate the travel times associated with the aforementioned velocity percentiles. We then track the evolution of the central galaxy associated with each halo using the SUBLINK merger trees created for TNG100, and collect the instantaneous energy injection rates from each feedback mechanism down to the earliest snapshot in TNG100. With the travel times and energy injection rate histories in hand, we can integrate from the three estimated launch times up to the snapshot in which we are measuring the gas properties. We determine the dominant feedback mechanism as the one whose integrated injected energy is highest for two out of the three estimated timescales.

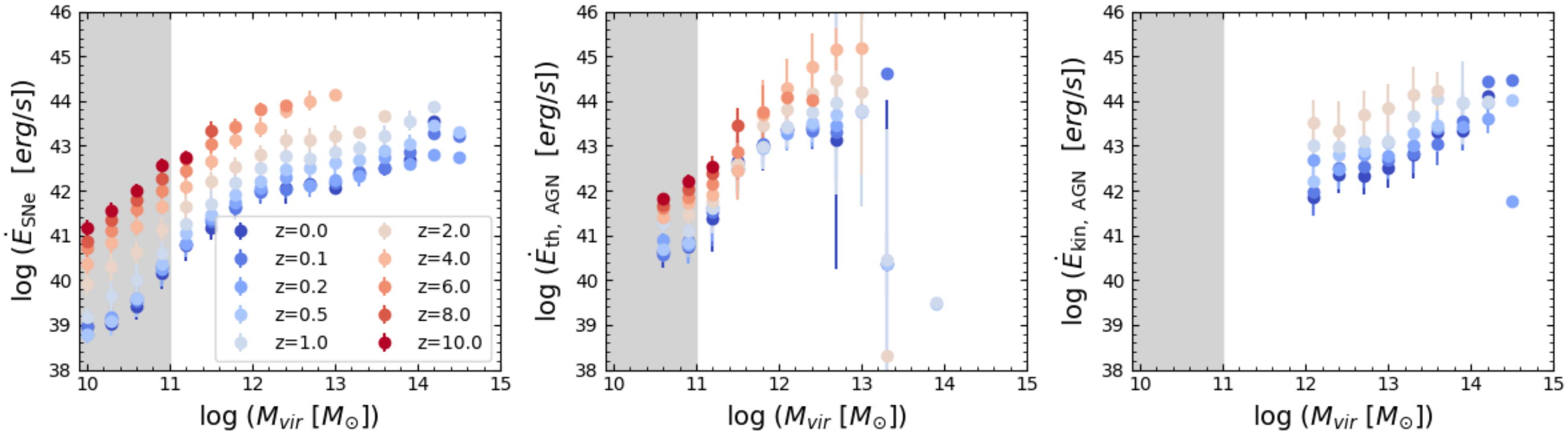

**Figure 3.** Instantaneous energy injection rates from the different feedback mechanisms in the TNG100 simulation, for all halos and redshifts analyzed in this work. Left panel: SN feedback, based on the subgrid recipe described in A. Pillepich et al. (2018a). Central panel: thermal-mode (i.e., high-accretion) AGN feedback. Right panel: kinetic-mode (i.e., low-accretion) AGN feedback. Both AGN energy injection rates are based on the subgrid recipe described in R. Weinberger et al. (2017). In all panels, the marker colors range from blue ($z = 0$) to red ($z = 10$), and the gray band marks the mass range whose associated central galaxies may be unresolved. SMBH generally begin their life in the thermal mode, and eventually transition to the kinetic mode.

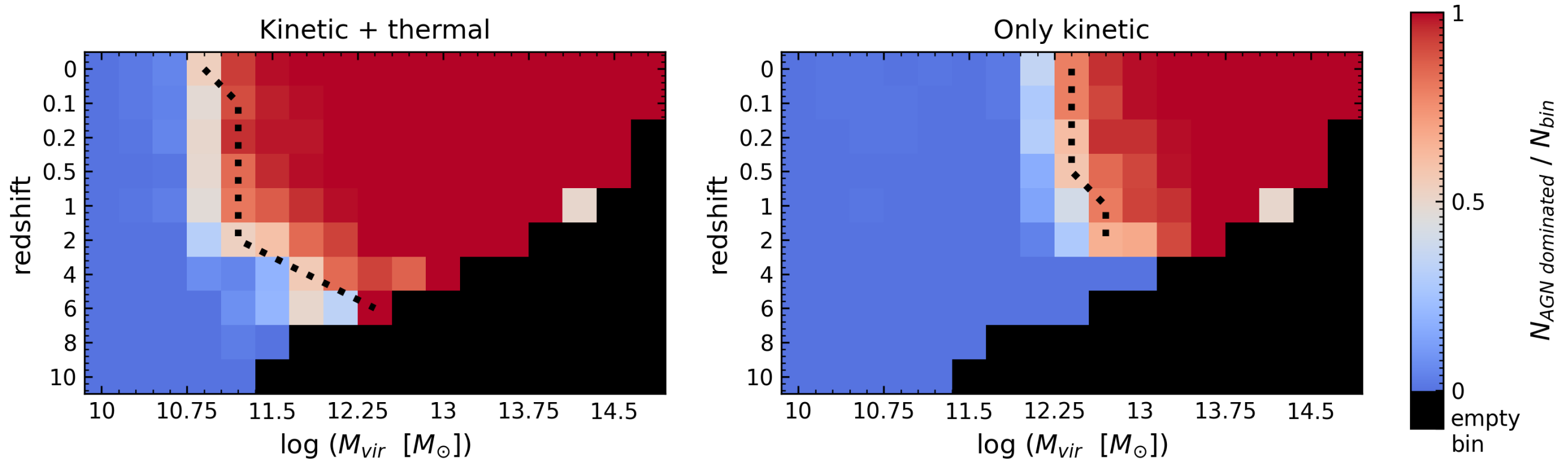

**Figure 4.** Fraction of halos, in redshift and virial mass bins, for which AGN are the dominant feedback energy injection mechanism. Fractions range from zero (blue—all halos in bin dominated by SN feedback) to one (red—all halos in bin dominated by AGN feedback). Bins with no halos are colored in black. We use a dotted black line to mark the lowest-mass bins, per redshift, where the fraction of AGN-dominated halos is greater than 0.5. Left panel: combination of kinetic + thermal feedback. Right panel: only kinetic feedback. In both cases, soon after AGN begin injecting energy into their surroundings, they become the dominant feedback energy injection mechanism.

Thermal feedback does not necessarily drive an outflow, as radiative losses have an appreciable effect (R. Weinberger et al. 2017). We therefore perform this comparison twice—once only for kinetic feedback, and once for kinetic and thermal feedback together (i.e., using the full feedback model).[9] The results are shown in Figure 4, in which we count the fraction of AGN-dominated halos in bins of virial mass and redshift.

We find that for $z \leqslant 6$, at some point there is a transition from SN-dominated feedback to AGN-dominated feedback as halos pass mass thresholds that decrease with cosmic time. Except for one case at $z = 1$, the transition is total, meaning that all halos become AGN dominated as we go to higher and higher masses.[10]

It is evident that soon after AGN feedback kicks in it becomes dominant, in both the kinetic case alone and for the full feedback model. As the energy injection rates plotted in Figure 3 show that thermal AGN feedback does not occur in halos more massive than $\sim 2 \times 10^{13}\,M_{\odot}$, the transitions between the dominant feedback energy injection mechanisms shown in Figure 4 are also an output of the "built-in" transition from thermal to kinetic AGN feedback in TNG100 halos. We hereafter refer to the virial mass bin in which, for a given redshift, the dominant feedback energy injection mechanism transitions from SN feedback to AGN feedback as "the onset of kinetic AGN feedback" when considering kinetic AGN feedback alone, and "the onset of thermal AGN feedback" when considering both thermal and kinetic modes. We use a dotted black line in Figure 4 to mark the onset of thermal (left panel) and kinetic (right panel) AGN feedback.

[9] While a BH particle in TNG100 can only be in either thermal mode or kinetic mode, a galaxy can contain more than one BH particle, and each of them can be in a different feedback mode.

[10] The outlier at $z = 1$ is the most massive halo for this snapshot in TNG100, and is one of only two halos in its mass bin (hence the fraction of AGN-dominated halos in the appropriate cell is exactly 0.5). This halo has a very high SFR—in fact, it is >6 times higher than the median SFR of halos in adjacent mass bins and the highest SFR in the entire $z = 1$ snapshot—while its BH accretion rate is close to that of similar mass halos.

## 4. Extracting the Flow Quantities

With an understanding of the dominant feedback energy injection mechanism for each virial mass bin and redshift, we can now analyze the gas flows in and out of galaxies and their host halos and correlate said flows with how the feedback is implemented. As mentioned in Section 2.1, for each of our 9522 halos we collect all gas, DM, and BH particles within $1.5R_{\rm vir}$. To estimate flow rates at the inner ISM scale, we select particles in shells between $0.05R_{\rm vir}$ and $0.15R_{\rm vir}$. At the halo scale the flow shells are from $0.95R_{\rm vir}$ to $1.05R_{\rm vir}$. To separate inflows from outflows, we consider the radial velocity component of each particle relative to the halo centers. Outflowing particles have positive radial velocities, and inflowing particles have negative radial velocities.

With each particle flagged by its direction and location, we can calculate the instantaneous mass inflow and outflow rates through a radial shell using

$$\dot{M} = \sum_i \dot{m}_i = \sum_i \frac{m_i v_{r,i}}{\Delta r}, \tag{9}$$

where the sum is over all outflowing or inflowing particles inside the shell, $\dot{m}_i$ is the mass flow rate for a single gas particle, $m_i$ is the particle mass, $v_{r,i}$ is its radial velocity, and $\Delta r = 0.1R_{\rm vir}$ is the width of the shell.

We consider two components for the energy flows. First is the thermal energy flow rate

$$\dot{E}_{\rm th} = \frac{3}{2}\sum_i \dot{m}_i c_{s,i}^2, \tag{10}$$

where $c_{s,i}^2$ is the particle sound speed. Second is the kinetic energy flow rate

$$\dot{E}_{\rm kin} = \frac{1}{2}\sum_i \dot{m}_i v_i^2, \tag{11}$$

where, here, $v_i$ is the total velocity of the gas particle (not just its radial component.) The total energy flow rate, kinetic plus thermal, is $\dot{E} \equiv \dot{E}_{\rm kin} + \dot{E}_{\rm th}$.

Finally is the metal mass flow rate

$$\dot{M}_Z = \sum_i \frac{m_i Z_i v_{r,i}}{\Delta r}, \tag{12}$$

where $Z_i$ is the heavy element mass fraction of the particle.

Similar to our discussion of resolved galaxies in TNG100, we may ask ourselves whether the flow rates themselves are resolved due to the small number of particles per shell. To inspect this, we tested how the number of particles in the ISM and halo shells scales with the halo virial mass, and whether the flow rates are sensitive to the shell width. We find that the number of gas particles per shell keeps a constant relation with the host halo mass down to our lowest-mass bin, and that changing our shell thickness from $0.05R_{\rm vir}$ to larger fractions of the virial radius does not alter any of our results at all halo masses and shell widths. These scalings alongside the large sample of 100 galaxies per mass bin lead us to conclude that our flow rates should be robust and properly represent the behavior in TNG100, regardless of whether the galaxy that drives the flow is resolved.

### 4.1. Loading Factors

As is standard, we define the loading factors relative to the SFRs, associated SN energy injection rates, and estimated metal injection rates. The mass loading factor is (as mentioned in Section 3.1)

$$\eta_M = \dot{M}_{\rm out}/{\rm SFR}, \tag{13}$$

where SFR is the instantaneous galaxy SFR as given by the simulation, and $\dot{M}_{\rm out}$ is our computed mass outflow rate.

The energy loading factor is

$$\eta_E = \dot{E}_{\rm out}/\left(\frac{\rm SFR}{100\,M_{\odot}}\right) \times E_{\rm SN}, \tag{14}$$

where $E_{\rm SN} = 10^{51}$ erg is the energy associated with an SN explosion, and ${\rm SFR}/(100\,M_{\odot})$ is the SN rate of $1\,M_{\odot}$ per $100\,M_{\odot}$ formed assuming a P. Kroupa (2001) IMF.

The metal loading factor is

$$\eta_Z = \dot{M}_{Z,{\rm out}}\ /\ (y \times {\rm SFR}). \tag{15}$$

Here $y = 0.02$ is the approximate metal yield from SNe, obtained using the same assumption for SN rate and a mean ejecta mass of $10\,M_{\odot}$, out of which 20% are metals (adopted by, e.g., C.-G. Kim et al. 2020a; V. Pandya et al. 2021).

Since in this work we define the loading factors relative to rates driven by star formation only (either the SFR itself, the energy released in SN explosions, or the rate of metals deposited into the ISM by SNe), we anticipate a significant increase in all loading factors at the onset of kinetic AGN feedback, or as we reach virial masses of $\sim 3 \times 10^{12}\,M_{\odot}$, above which galaxies begin to quench (see Figures 4 and 2) and AGN become the dominant forces in driving gas, energy, and metals out of the ISM. Therefore, in this work we only present the loading factors for halos with virial masses of less than $3 \times 10^{12}\,M_{\odot}$.

## 5. Mass Flow Rates

We now start with analysis of the mass flow rates through the inner galaxy ISM shells and outer halo shells. We begin with a straightforward analysis of inflow and outflow rates, and whether galaxies and halos gain or lose mass. We touch on the phenomena behind these net flows, and how they are correlated with other, using more commonly discussed properties of TNG100 halos and galaxies (SFR and baryon fractions). We then focus on the mass inflow rates, first in the context of their deviation from expectations based on DM mass flow rates, and later we track their progression through the halo scale, down to the ISM scale and as they become fuel for star formation. We end with the mass loading factors out of TNG100 galaxies.

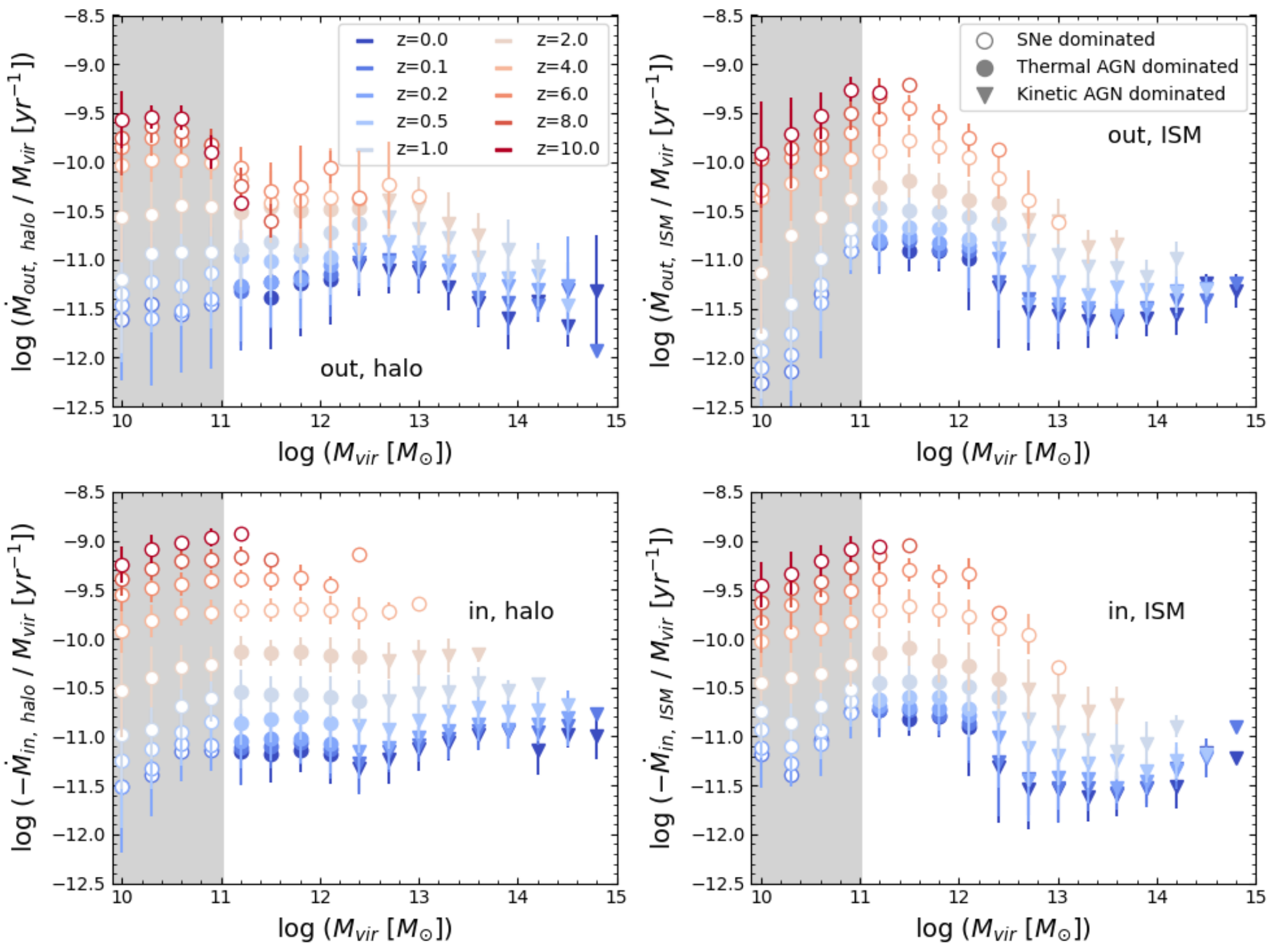

**Figure 5.** Gas mass outflow rates (top row) and inflow rates (bottom row), estimated at the halo scale (left column) and at the ISM scale (right column), all normalized by the halo virial mass and presented as a function of the halo virial mass. The gas mass flow rates are calculated from $z = 10$ (red markers with error bars) to $z = 0$ (blue markers with error bars). For each redshift, we mark halos dominated by SN feedback with open circles, halos dominated by thermal AGN feedback with full circles, and halos dominated by kinetic AGN feedback with full triangles.

### *5.1. Net Inflows and Outflows*

In Figure 5 we plot the gas mass inflow and outflow rates for all redshifts at the ISM and halo scales, as functions of the virial mass. We normalize by $M_{\rm vir}$ since to first order we expect the baryonic inflow rates to be proportional to the DM inflow rates, which are themselves approximately proportional to the total virial masses (e.g., S. Genel et al. 2008). For each redshift, we mark mass bins dominated by SN feedback with open circles, mass bins dominated by thermal AGN feedback with full circles, and mass bins dominated by kinetic AGN feedback with full triangles. In Figure 6 we plot the ratios of the inflow to outflow rates, at the ISM and halo scales, to determine whether there are net inflows or outflows at these two scales. The colors and markers are as in Figure 5.

Several trends appear in Figures 5 and 6. At the halo scale (left panels of Figure 5) we find different trends for inflows and for outflows. For low virial masses and at high redshifts ($z \geqslant 4$), the normalized outflow rates are relatively constant up to halo virial masses of $\sim 10^{11}\,M_\odot$, after which they begin decreasing. At lower redshifts, however, the normalized mass outflow rates increase with the halo virial mass. Above $3 \times 10^{12}\,M_\odot$, the normalized mass outflow rates decrease at all redshifts, regardless of the feedback mechanism (the absolute outflow rates are still increasing). The normalized mass inflow rates, on the other hand, slightly increase with the halo virial mass for all redshifts up to $\sim 3 \times 10^{11}\,M_\odot$, after which they decrease until reaching the onset of kinetic AGN feedback, and then increase again. As seen in Figure 6 the net flows are predominantly inward by large factors. The exceptions are low-redshift ($z \leqslant 0.5$) halos near $\sim 3 \times 10^{12}\,M_\odot$ at the onset of kinetic AGN feedback where substantial net outflows occur. This is consistent with the results shown in, e.g., J. J. Davies et al. (2020), R. J. Wright et al. (2024), and Y. Oren et al. (2024), which show a minimum of the CGM gas mass fraction at this virial mass for $z = 0$. An additional case of net mass outflows at the halo scale is seen for dwarf galaxies (at the centers of $\sim 10^{10}\,M_\odot$ halos) at $z = 0$, where we find slightly higher outflow rates than inflow rates. Unlike for the onset of kinetic AGN feedback, we do not find these net outflows at the ISM shells surrounding dwarf galaxies; in fact, at this mass bin we find ISM inflows to be the highest relative to ISM outflows.

Transitioning to the ISM scale (right panels), Figure 5 shows that the inflow and outflow rates are comparable. For a given redshift, the normalized flow rates at the ISM shell (in both directions) increase with mass and peak near $\sim 2 \times 10^{11}\,M_\odot$. They then decline, and increase again for halo virial masses of $\sim 2 \times 10^{13}\,M_\odot$. As seen in Figure 6 (right panel) overall the net flows are into the ISM, enabling the galaxies to grow. The net inflow rates decrease with increasing halo mass. At the particular redshifts and masses where kinetic AGN feedback turns on, the outflow rates become nearly equal to the inflow rates, with a slight preference toward net outflows. At redshifts where kinetic AGN feedback is not active, the flows remain inward for all virial masses. The balance between the galactic outflow and inflow rates persists up to halo virial masses $\sim 10^{14}\,M_\odot$, after which the ISM inflow rates again become significantly larger than the outflow rates—most likely due to the growing binding energy of the galaxies within. In low-mass halos the ratios of the inflow to outflow rates can reach rather large values (>0.25 dex). Strong net galactic inflows also appear in high-mass ($M_{\rm vir} > 10^{12}\,M_\odot$), high-redshift ($z \geqslant 4$) halos in which there is no recent AGN feedback.

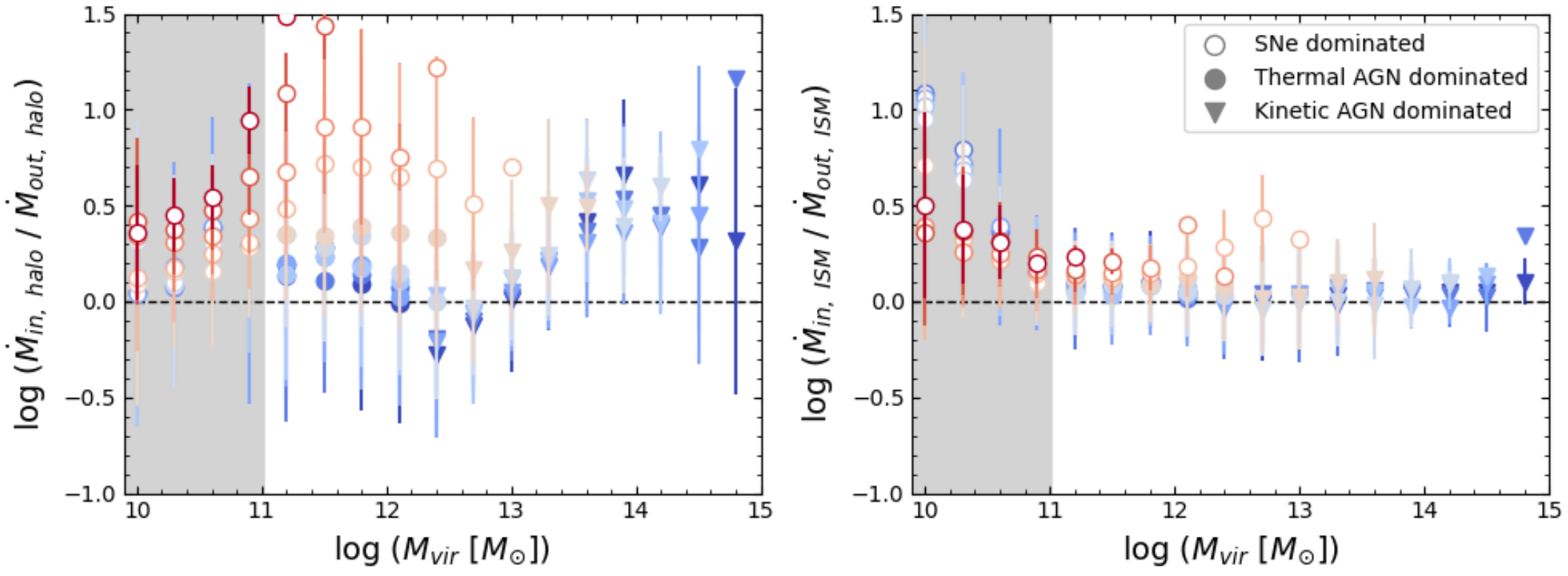

**Figure 6.** The ratio between the gas mass inflow and outflow rates at the virial scale (left panel) and the ISM scale (right panel). Colors and symbol types are identical to those in Figure 5. In both panels, and for most redshifts and virial masses, $\dot{M}_{\rm in} > \dot{M}_{\rm out}$, suggesting a net inflow into both halos and galaxies. The net inflows, especially on halo scales, are much stronger at high redshift.

To further investigate the correlation between the onset of kinetic AGN feedback and the decrease in the baryon fractions of TNG100 halos, we turn to Figure 7. In it, we show the fractions of the virial mass that are taken by the total baryonic mass within the virial radius, and those taken by the CGM mass, both normalized by the cosmological baryon fractions (labeled $f_{\rm Bar,halo}$ and $f_{\rm CGM}$, respectively). At redshifts where kinetic AGN feedback is active, there is a local minimum of both $f_{\rm CGM}$ and of $f_{\rm Bar,halo}$ around the onset of kinetic AGN feedback, similar to the case at $z = 0$ and in accordance with the appearance of net mass outflows at the halo scale. The drop in both the CGM and baryon factions is more significant at lower redshifts, where presumably kinetic AGN feedback has been active longer. On the other hand, at $z = 4$, where there is no kinetic AGN feedback, $f_{\rm CGM}$ decreases past $M_{\rm vir} \approx 10^{12}\,M_\odot$ while the total baryon fraction remains relatively constant. This suggests that at this redshift, the baryons lost from the CGM fall into the galaxy, unlike at lower redshifts where they are ejected out of the halo entirely due to AGN feedback. This is supported by the right panel of Figure 6, showing a strong net inflow into the galaxy at high masses at $z = 4$, and the previously discussed increase in SFRs for this case. At higher redshifts our sample does not include enough massive halos to investigate this trend.

### 5.2. Comparison with Dark Matter Halo Growth Rates

To first order, we expect the rate that baryons flow into a halo to be proportional to the rate that the halo is accreting DM. We examine deviations from this assumption in Figure 8, in which we present the gas mass inflow rates normalized by $f_{\rm B}\dot{M}^{\rm DM}_{\rm in,halo}$, the DM mass inflow rate at the virial radius multiplied by the cosmological baryon fraction. We show these at the halo scale (left panel) and the ISM scale (right panel).

At the halo scale the normalized flow rates are usually close to but somewhat smaller than unity. We find two interesting cases for which the inflows are significantly reduced: the first is for dwarf galaxies at all redshifts, and the second is at masses associated with the onset of kinetic AGN feedback.

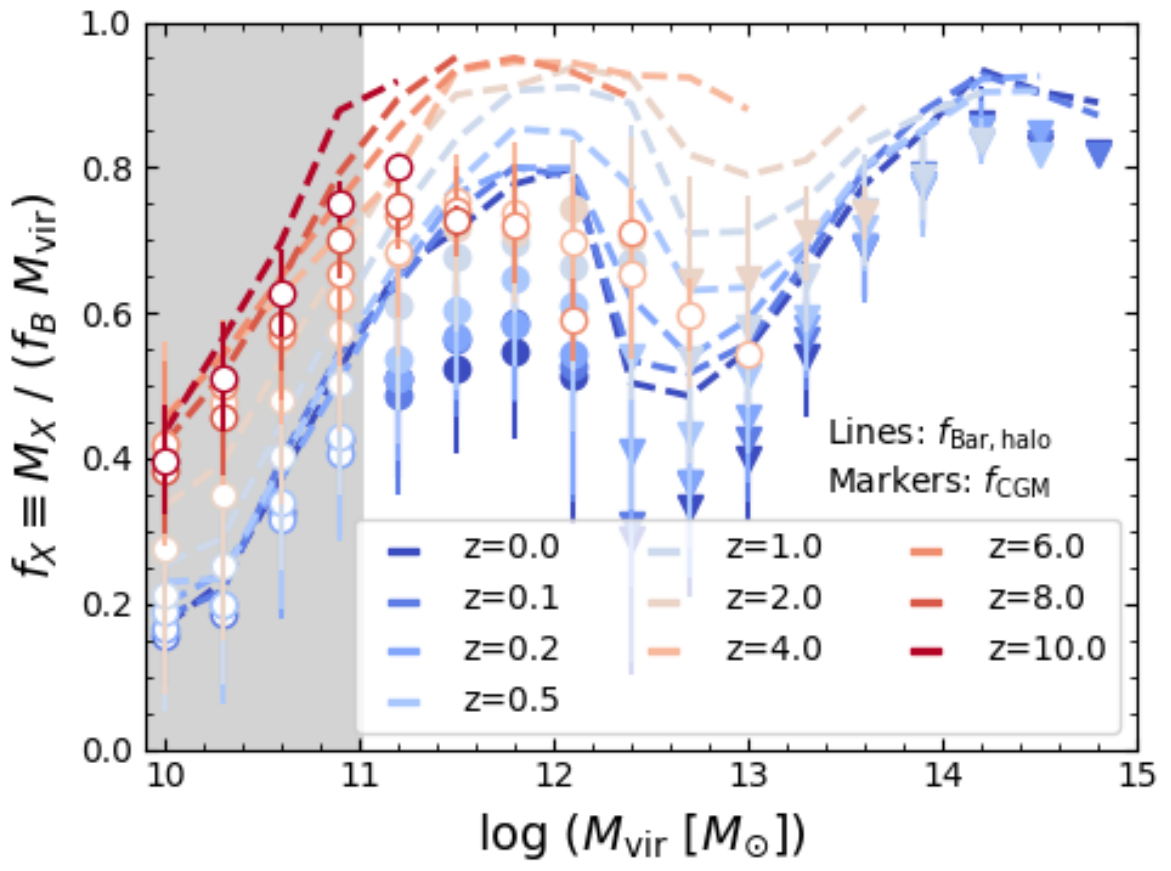

**Figure 7.** Baryon fractions, defined as the combined mass of all baryons within a certain population normalized by $f_{\rm B}M_{\rm vir}$, for the CGM alone (scattered markers with error bars) and for the entire halo (dashed lines). Colors and marker styles are identical to those in Figure 5. At redshifts where kinetic AGN feedback is active, both $f_{\rm CGM}$ and $f_{\rm Bar,halo}$ decrease at the onset of kinetic AGN feedback. At redshifts where there is no kinetic AGN feedback, the drops in $f_{\rm CGM}$ and $f_{\rm Bar,halo}$ are less significant.

Both reductions are consistent with strong feedback suppressing inflows into the galaxy; in the case of the low-mass halos, the $z = 0$ point (numerically the lowest) is also a case where there are net outflows from the halo, but not the galaxy as seen in Figure 6. The case of halos at the onset of kinetic AGN feedback is also consistent with Figure 6, where we find net outflows at these masses. An additional interesting outlier at the halo scale are high-redshift ($z \geqslant 2$) inflows into halos with virial masses $\gtrsim 10^{11}\,M_\odot$, which are higher than expected based on the DM flow rates alone, suggesting recycling of previously ejected gas at these scales.

At the ISM scale in Figure 8 we find two notable features. First, at virial masses around $10^{11}\,M_\odot$ the normalized mass inflow rates exceed unity by up to 0.25 dex. This excess may also suggest recycling of galactic outflows that are falling back into the galaxy (also known as a "galactic fountain"), in addition to fresh IGM gas streaming into the halo. Second, in dwarfs and in more massive halos ($M_{\rm vir} \gtrsim 3 \times 10^{12}\,M_\odot$) the baryonic mass inflow rates are reduced at the ISM scale. This is a signature of preventative feedback: fresh IGM gas enters the halo but is prevented from reaching the ISM, contributing to growth of the CGM (see, e.g., R. Davé et al. 2012). This preventative feedback is caused by SN-driven winds at the low-mass end, and by kinetic AGN feedback at the high-mass end. We measure the strongest preventative feedback in halos with $M_{\rm vir} \approx 2 \times 10^{13}\ M_\odot$, above which the galactic mass inflow rates become closer to the expectations from DM flows. Still, galaxies in TNG100 do not reach masses which allow their potential wells to overcome the energy injected by their AGN.

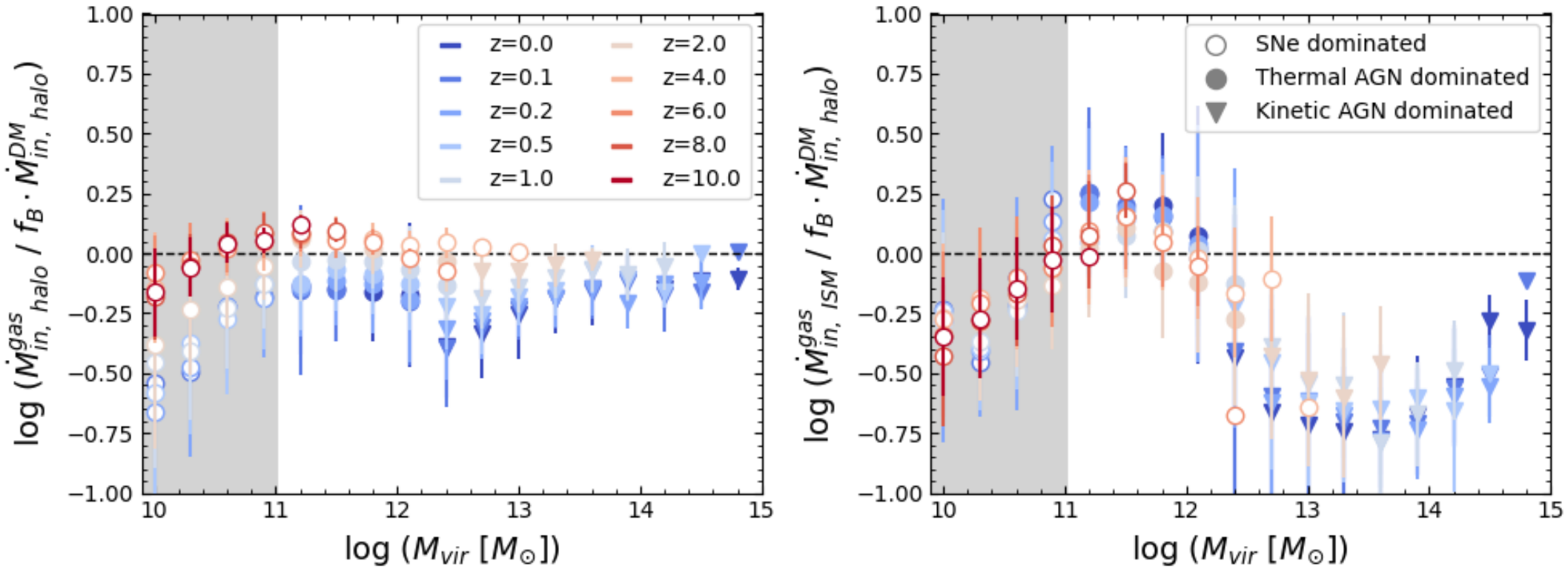

**Figure 8.** Gas mass inflow rates, estimated at the halo scale (left panel) and at the ISM scale (right panel), normalized by the DM inflow rates multiplied by the baryon fraction. Colors and symbol types are identical to those in Figure 5. At the halo scale, we see inflow rates that slightly exceed the baseline rate $f_B\,\dot{M}^{\rm DM}_{\rm in,halo}$ at high redshift, and are lower than the baseline at low redshift, indicating preventative feedback. At the ISM scale, at halo masses $\sim 10^{11}$–$10^{12}\ M_\odot$, we see inflows in excess of the baseline, indicating rapid recycling of ejected gas on this scale in a galactic fountain. At higher halo masses we see a strong suppression of inflows due to preventative feedback, primarily caused by the kinetic-mode AGN feedback.

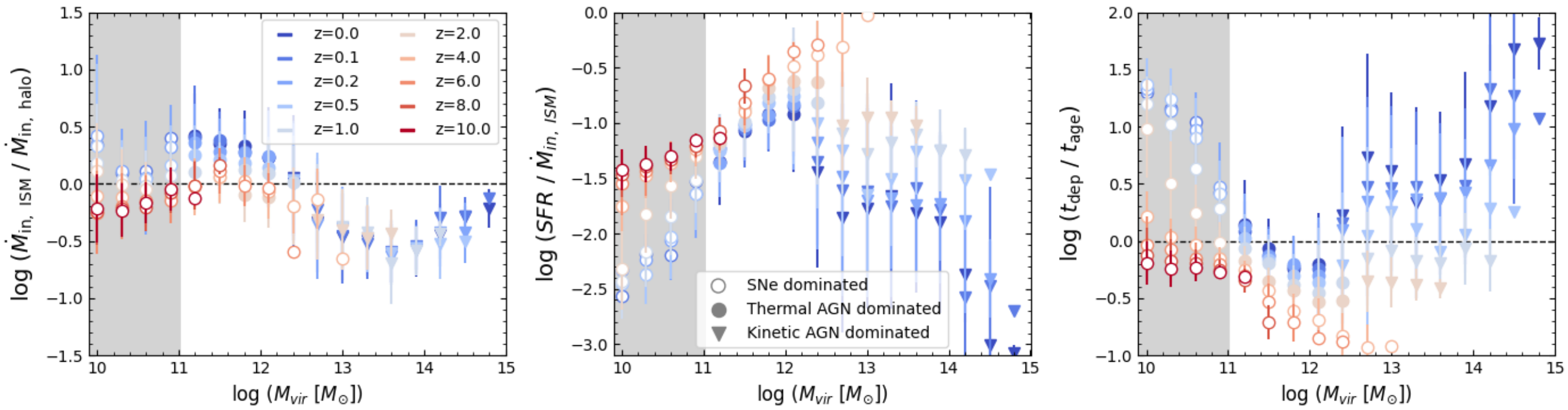

**Figure 9.** Left panel: the ratio between the mass inflow rates at the ISM scale to those at the halo scale, indicating which fraction of the gas flowing into the halo makes it to the galaxy. At virial masses of $\sim 3 \times 10^{11}\ M_\odot$ we find that fraction to be larger than one, indicating rapid cooling and infall of previously ejected ISM gas in addition to fresh IGM gas; in halos with virial masses of $\gtrsim 10^{12}\ M_\odot$, we find an indication of preventative feedback and the buildup of a hot CGM. Central panel: the ratio between the gas mass flow rates into the galaxy and the instantaneous SFR in the central galaxy, indicating which fraction of the galactic inflows was used for star formation. This fraction initially increases with halo mass, only to drop after the onset of kinetic AGN feedback. Right panel: $t_{\rm dep}$, the ISM depletion time, normalized by the age of the universe ($t_{\rm age}$). For all redshifts, the depletion time drops below the age of the universe in halos with virial masses of $\sim 10^{11}$–$10^{12}\ M_\odot$. The depletion time becomes long compared to $t_{\rm age}$ in more massive halos. In all panels, the colors and symbol types match those of Figure 5.

### 5.3. Progression of Inflows

In Figure 9 we address the questions of what fraction of the gas mass that flows into a halo enters the galaxy (left panel), how much of the entering gas is used for star formation (central panel), and with what efficiency the galaxy converts accreted gas into stars (right panel).

For most virial masses and redshifts, the left panel of Figure 9 is consistent with the comparison between the baryonic inflow rates to the DM inflow rates. For virial masses above $\sim 3 \times 10^{12}\ M_\odot$ at all redshifts, and for for virial masses lower than $\sim 10^{11}\ M_\odot$ at $z \geqslant 2$, the halo inflow rates are higher than the ISM inflow rates, consistent with the markers of preventative feedback. For halos at $z \leqslant 2$ with virial masses $\sim 10^{11}\ M_\odot$ the inflow rates into the galaxy surpass those that fall into the halo, again supporting an additional contribution to the gas falling in from the IGM, perhaps in the form of previously ejected gas cooling and falling back into the galaxy (a "galactic fountain").

Interesting outliers, however, are halos in the virial mass bin of $M_{\rm vir} = 10^{10}\ M_\odot$. At $z \leqslant 2$, we find that inflows into the ISM surpass those into the halo, while a comparison with the DM flow rates suggested reduced inflows into both the halo and the ISM at these masses and times. Recall that at $z = 0$ we find net outflows from the halo scale at this mass bin, while simultaneously seeing strong net inflows into the ISM.

The central panel of Figure 9 shows the ratio between the SFRs of the central galaxies to the inflow rates into the ISM. For halos with $z \leqslant 2$, the fraction of inflowing gas that is used for star formation increases with virial mass, only to decrease after the onset of kinetic AGN feedback. For $z \geqslant 4$, where kinetic AGN feedback has not yet occurred, there is no turnover and the ratio keeps increasing with virial mass. Most notably, for the highest-mass bin at $z = 4$, all gas that flowed into the galaxy is used for star formation. While at $z \leqslant 2$ thermal AGN feedback appears to be in effect, it is not as effective at suppressing star formation as kinetic AGN feedback: the dependence of $\mathrm{SFR}/\dot{M}_{\mathrm{in,ISM}}$ on the halo virial mass past the onset of thermal AGN feedback does not appear different from that of halos where SN feedback is the dominant feedback mechanism.

The right panel of Figure 9 shows the ratio between the ISM depletion time, defined as $t_{\mathrm{dep}} \equiv M_{\mathrm{ISM}}/\mathrm{SFR}$, normalized by the age of the universe $t_{\mathrm{age}}$. By comparing the two timescales, we can get a sense of the star formation efficiency within the galaxy. In low-mass halos, with virial masses $\lesssim 10^{11}\,M_{\odot}$, the depletion times are very high at recent times, yet at $z \geqslant 4$ they converge to 60% of the age of the universe. This suggests SN feedback at these masses becomes less effective in suppressing star formation as time progresses. On the other hand, star formation is most effective in galaxies populating halos between $\sim 10^{11}\,M_{\odot}$ and $\sim 3 \times 10^{12}\,M_{\odot}$, regardless of the dominant feedback mechanism—SN feedback or thermal AGN feedback.

In halos more massive than $\sim 3 \times 10^{12}\,M_{\odot}$, the behavior varies depending on the dominant feedback mechanism. At the onset of kinetic AGN feedback we measure an increase of the depletion time, which becomes more drastic the lower the redshift. At $z = 2$ and $z = 1$, when kinetic AGN feedback initially kicks in, it slightly increases the depletion time but still does not bring it above the age of the universe. At $z \leqslant 0.5$, however, the SFRs decrease rapidly as kinetic AGN feedback drives significant amounts of gas out of the galaxies. On the other hand, at redshifts where kinetic AGN feedback is not activated, we find that the depletion time keeps decreasing, showing increasingly more intense star formation and insufficient feedback to restrain it.

### 5.4. Mass Loading Factors

Before discussing the loading factors out of the TNG100 galaxies, we should justify our claim that past the onset of kinetic AGN feedback, the loading factors become relatively uninformative—at least in the context in which they are defined in this work. In the left panel of Figure 10 we plot the ratio between $\eta_M$, the mass loading factor we define in Equation (13), and $\eta_{w,\,\mathrm{TNG}}$, the input TNG100 mass loading factor defined in Equation (2). Until the onset of kinetic AGN feedback, the ratio between $\eta_M$ and $\eta_{w,\,\mathrm{TNG}}$ is close to unity (ranging from a ratio of $\sim$0.3 for our lowest-mass bin at $z = 10$ to $\sim$5 for $M_{\mathrm{vir}} \approx 3 \times 10^{10}\,M_{\odot}$ at $z = 0$). This agreement also continues for masses above our suggested mass cut of $3 \times 10^{12}\,M_{\odot}$—but only at redshifts where AGN feedback is not yet active.

On the other hand, once kinetic AGN feedback becomes the dominant feedback mechanism, the ratio between the measured and the TNG input mass loading factors increases dramatically—up to a ratio larger than 1000 for the highest-mass bin at $z = 0$. Since $\eta_{w,\,\mathrm{TNG}}$ governs the rate at which wind particles are launched in TNG, this suggests that past the onset of kinetic AGN feedback, SNe play a negligible role in driving outflows. Notably, $\eta_M/\eta_{w,\,\mathrm{TNG}}$ remains within $O(1)$ for halos dominated by thermal AGN feedback, suggesting it is subdominant in driving outflows. We also note that for all virial masses, the ratio between $\eta_M$ and $\eta_{w,\,\mathrm{TNG}}$ decreases with redshift, such that at high redshifts, less material emerges in the wind than has been launched, suggesting that the winds may have stalled, while at low redshifts, the mass flow rates are higher than they were when they were launched. The independence of this result on the halo virial mass leads us to conclude that this phenomenon is related to the higher density of the CGM at early times.

The central panel of Figure 10 shows the measured mass loading factors at the ISM scale, as described in Equation (13). There are strong trends with both halo mass and redshifts. For $z \geqslant 4$, the mass loading factors are roughly 10 for all mass bins lower than $10^{11}\,M_{\odot}$, and decreases monotonically toward higher masses. The mass loading factors increase with decreasing redshifts but keep a relatively similar functional form, as they remain relatively flat for $M_{\mathrm{vir}} \lesssim 3 \times 10^{10}\,M_{\odot}$ and then also monotonically decrease with higher masses. At $z = 0$, we reach $\eta_M \approx 100$ at the lower-mass end. In fact, for all halo mass bins the loading factors become higher as the redshift decreases, and the difference between $z = 0$ and $z = 10$ is largest in the $10^{10}\,M_{\odot}$ mass bin (this statement is true for both $\eta_M$ and $\eta_M/\eta_{w,\,\mathrm{TNG}}$). For $z \leqslant 2$, the mass loading factors decrease with virial mass. Just before the increase at the $\log(M_{\mathrm{vir}}/M_{\odot}) = 12.1$ mass bin, $\eta_M$ drops to $\sim$6 for $z = 0$ and $\sim$4 for $z = 2$. At redshifts where kinetic AGN feedback does not kick in, the mass loading factor keeps decreasing with redshift until reaching 1–2 in the $\log(M_{\mathrm{vir}}/M_{\odot}) = 12.5$ mass bin. The onset of thermal AGN feedback does not seem to affect the mass loading factors: for example, the dependence of $\eta_M$ on the halo virial mass at $z = 0$ does not change with the onset of thermal AGN feedback in the $\log(M_{\mathrm{vir}}/M_{\odot}) = 11.1$ mass bin, while the onset of kinetic AGN feedback in the $\log(M_{\mathrm{vir}}/M_{\odot}) = 12.4$ mass bin increases it significantly.

The right panel of Figure 10 shows the ratio between the mass loading factor at the halo scale and the mass loading factor at the ISM scale (equivalent to $\dot{M}_{\mathrm{out,halo}}/\dot{M}_{\mathrm{out,ISM}}$ for galaxies with $\mathrm{SFR} > 0$). For halos with virial masses $< 3 \times 10^{10}\,M_{\odot}$, the mass loading factors increase as they travel through the CGM. For higher virial masses the loading factors decrease by up to $\sim$0.5 dex, with no apparent dependence on redshift—except for $z = 8$ and $z = 10$ where the loading factors significantly decrease by $\sim$1 dex. The drop is strongest for virial masses of $\sim 3 \times 10^{11}\,M_{\odot}$, after which the loading factor ratios start increasing until the onset of kinetic AGN feedback. It is notable that the trend shown by the ratio of loading factors (or outflow rates) is the opposite of the trend shown by the ratio of inflow rates, as seen in the left panel of Figure 8. There is a correlation between a decrease (increase) in inflows from the halo shell to the ISM shell, and an increase (decrease) in outflows from the ISM shell to the halo shell.

## 6. Energy Flow Rates

In this section we measure flows of energy in TNG100. We start with the energy inflow and outflow rates in all directions and scales, in addition to discussing whether energy flows into or out of TNG100 galaxies and halos. We then continue with energy loading factors.

### 6.1. Net Inflows and Outflows

In Figure 11 we plot the energy flow rates for all of our selected halos. In as much as the baryons follow the DM, we expect the gaseous energy flow rates to be proportional to the DM energy flow rates, $\sim\frac{1}{2}\dot{M}_{\mathrm{DM}} V_{\mathrm{vir}}^2$, where $V_{\mathrm{vir}}$ is the virial velocity of the parent halo. Hence, we normalize the energy flow rates by $M_{\mathrm{vir}} V_{\mathrm{vir}}^2 = G M_{\mathrm{vir}}^2/R_{\mathrm{vir}}$. As in Figure 5 for the mass flows, in Figure 11 we plot the normalized energy inflow and outflow rates at the halo and ISM scales. In Figure 12 we plot the ratios of the energy inflow to outflow rates.

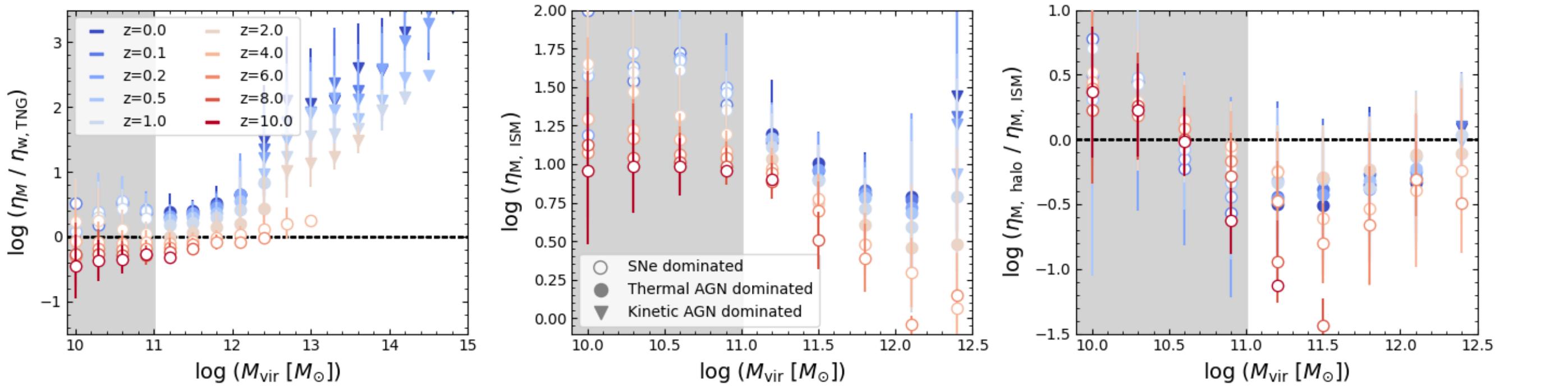

**Figure 10.** Left panel: the ratio between the mass loading at the ISM scale (estimated as $\dot{M}_{\rm out,ISM}/{\rm SFR}$) and the mass loading of injected wind material in the TNG subgrid recipe described in A. Pillepich et al. (2018a). This ratio is close to unity or slightly below it at high redshift, and increases with decreasing redshift. Central panel: mass loading factors at the ISM scale for halos where SNe are the dominant feedback mechanism. Generally, loading factors at a given halo mass decrease with redshift, to the point where at $z = 10$ they may be 1 dex lower than at the present day. For a given redshift, as the virial masses increase and with them the halo gravitational potentials, the loading factors decrease. Right panel: the ratio between the mass loading factors at the halo scale and the mass loading factors at the ISM scale. Mass outflow rates decrease as they travel through the CGM, except for low-mass dwarfs and halos at the onset of kinetic AGN feedback. For all panels, the colors and symbol types match those of Figure 5.

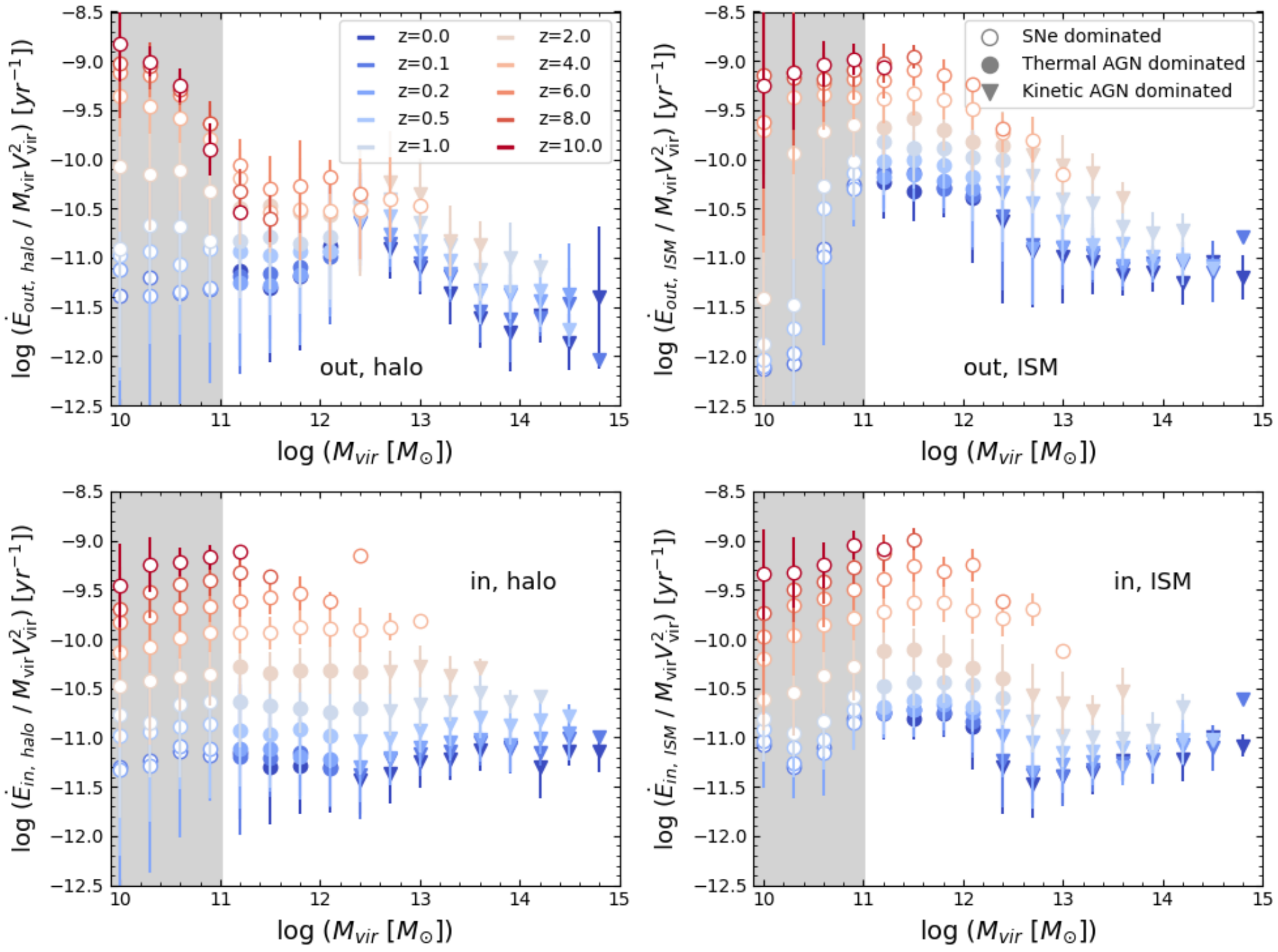

**Figure 11.** Gas energy outflow rates (top row) and inflow rates (bottom row), estimated at the virial radius (left column) and at the ISM radius (right column), all presented as a function of the halo virial mass and normalized by $M_{\rm vir}V_{\rm vir}^2 = GM^2/R$. Colors and symbol types match those of Figure 5.

The normalized energy flow rates are very similar to the normalized mass flow rates shown in Figure 5, at both the halo and ISM scales. Most notably, both the normalized energy flow rates and normalized mass flow rates occupy the same dynamic range, and show a similar dependence on both halo virial mass and redshift.

At the halo scale (left panels of Figure 11), we find no significant differences between the normalized energy inflow rates and their mass counterparts: for a given redshift, the normalized inflow rates slightly increase with halo virial mass, except for a dip at the onset of kinetic AGN feedback. There are, however, two notable differences between the normalized energy outflow rates and the normalized mass outflow rates. First, the high-redshift ($z \geqslant 4$) normalized energy outflow rates increase with decreasing halo virial mass, while the normalized mass outflow rates at the same redshifts flatten for $M_{\rm vir} < 10^{11}\,M_\odot$. Second, the normalized energy outflow rates from halos with virial masses between $10^{12}\,M_\odot$ and $10^{13}\,M_\odot$ at $z \leqslant 2$ appear to be relatively higher than the normalized mass flow rates at the same mass bin. The normalized energy outflow rates at other masses and redshifts do not differ from their mass counterparts.

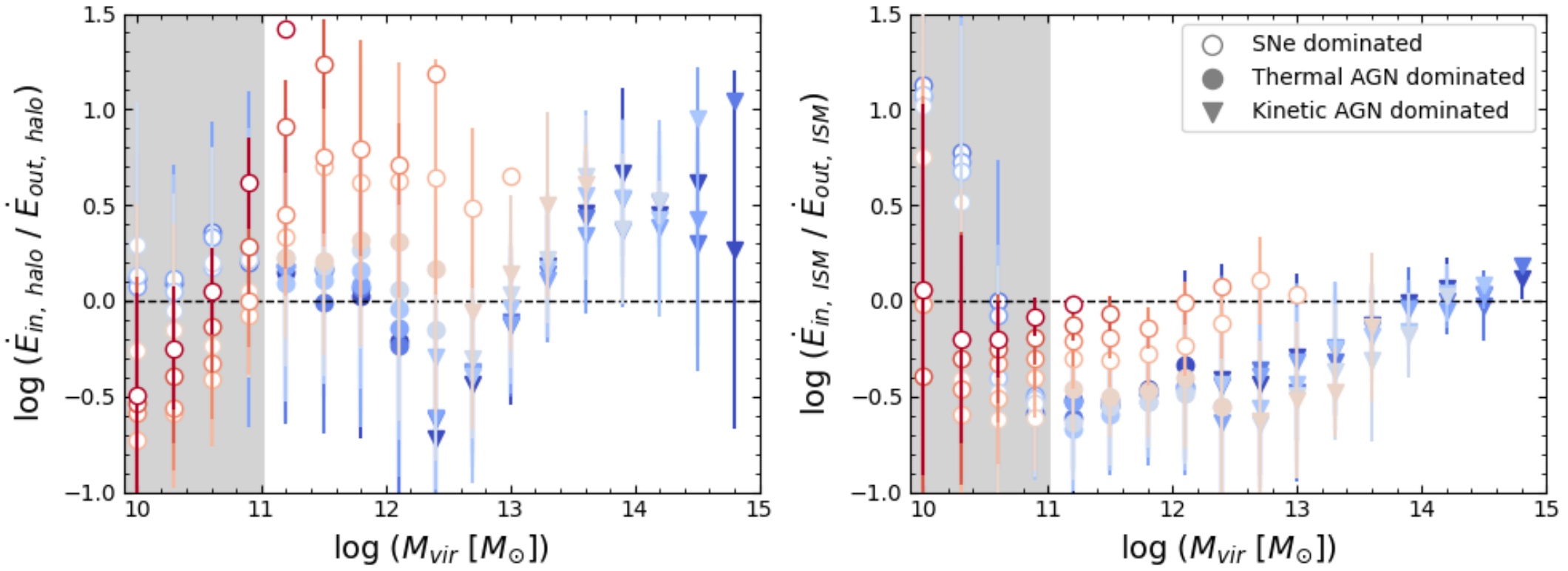

**Figure 12.** The ratio between the energy inflow and outflow rates at the halo scale (left panel) and the ISM scale (right panel). Colors and symbol types match those of Figure 5. At the halo scale there is a net outflow of energy only right around the mass scale where kinetic-mode AGN feedback becomes active. At the ISM scale, there is a net outflow of energy for almost all virial masses and redshifts.

The two differences between the mass and energy flow rates at the halo scale also manifest themselves as we observe the ratio between the energy inflows and outflows at this scale in Figure 12. For most redshifts and mass bins, there is a net inflow of energy into the halos. However, at $z \leqslant 1$ there is a very significant net energy outflow from the halo scale at virial masses between $10^{12}\,M_\odot$ and $10^{13}\,M_\odot$. While all halos that show these energy outflows are dominated by AGN feedback, a significant fraction of those energy outflows occur in halos whose dominant feedback mechanism is thermal AGN feedback. This may either be SN-driven outflows that are also heated by thermal AGN feedback, or kinetic AGN feedback that occurs in some of these halos and is strong enough to raise the median energy outflow rate above the median energy inflow rate (the fraction of kinetic AGN-dominated halos in this mass bin is lower than 0.5—which is why the bin itself is not considered dominated by kinetic AGN feedback—but the fraction is still larger than zero; see right panel of Figure 4). Another case where we measure a net outflow of energy from the halo scale is for high-redshift dwarfs, at $z \geqslant 4$ and with $M_{\rm vir} < 10^{11}\,M_\odot$, which becomes more significant at lower masses and earlier times. These outflows must be driven by SNe, which do not lead to mass outflows from the halo (as mass flows inward) but cause preventative feedback. For all other mass bins and redshifts, energy is flowing into the CGM from both the galaxy and the IGM.

Transitioning to the ISM scale (right panels of Figure 11), we find one significant difference between mass and energy flow rates: past the onset of kinetic AGN feedback, both normalized mass outflows and normalized energy outflows decrease. While the normalized energy outflow rates continue to decrease with virial mass until reaching the highest-mass bin, the normalized mass outflow rates increase again past halo virial masses of $\sim 2 \times 10^{13}\,M_\odot$.[11] While the normalized galactic energy outflow rates decrease, the normalized galactic energy inflow rates begin to increase past the onset of kinetic AGN feedback. As we compare the two in Figure 12, we find that galactic inflow rates surpass the galactic outflow rates only in halos whose virial masses are higher than $\sim 10^{14}\,M_\odot$.

[11] This transitional mass bin is where we measure the strongest preventative feedback caused by AGN (as discussed in Section 5.2).

As a matter of fact, the right panel of Figure 12 shows that the net energy flows are outward at most masses and redshifts, whereas the net mass flows are inward. There are net energy inflows into TNG100 galaxies only in three regimes: in low-redshift dwarfs, where we also find very strong mass inflows into the galaxy alongside net outflows from the halo; at the most massive end of the $z = 4$ sample, where no AGN feedback is yet active and the net mass inflow rates are strong; and at masses above $M_{\rm vir} \approx 10^{14}\,M_\odot$. For all other masses and redshifts, energy flows outward from galaxies while they gain mass.

### 6.2. Energy Loading Factors

In Figure 13 we plot the energy loading factors of the central TNG100 galaxies in our sample, calculated using Equation (14), with markers and colors following our usual convention. The top panel shows the energy loading factors at the ISM shell, and the bottom panel shows the ratio between the energy loading factors at the halo shell to those at the ISM shell (equivalent to plotting $\dot{E}_{\rm out,halo}/\dot{E}_{\rm out,ISM}$ for halos with SFR $> 0$).

Before the onset of kinetic AGN feedback, the energy loading factors from the ISM of TNG100 galaxies range between $\sim$0.5 (for $z = 0$ halos with $M_{\rm vir} \approx 3 \times 10^{11}\,M_\odot$) and $\sim$1.5 (for $M_{\rm vir} \approx 4 \times 10^{10}\,M_\odot$ halos at $z = 2$). This narrow range of energy loading factors is independent of redshift or halo mass. However, the two lowest-mass bins show a pronounced decrease of energy loading factors at later times. For example, for $z = 0$ $\eta_M$ increases by $\sim$0.9 dex between the second ($2 \times 10^{10}\,M_\odot$) and third ($4 \times 10^{10}\,M_\odot$) mass bin, while for $z = 10$ it only increases by a factor of $\sim$0.2 dex. The onset of thermal AGN feedback at $M_{\rm vir} \approx 10^{11}\,M_\odot$ has no effect on the loading factors.

While the mass loading factors at the ISM shell are very different than the corresponding energy loading factors, the evolution of the two loading factors from the ISM shell to the halo shell is remarkably similar. For dwarfs $\eta_{\rm E,halo} > \eta_{\rm E,ISM}$, but as the halo masses increase the distance between loading factors quickly narrows and then flips until $\eta_{\rm E,halo} \approx 0.1\eta_{\rm E,ISM}$ for $M_{\rm vir} \approx 3 \times 10^{11}\,M_\odot$. At higher halo masses the energy loading factor at the halo shell again increases relative to the ISM shell. For all cases there is no redshift dependence.

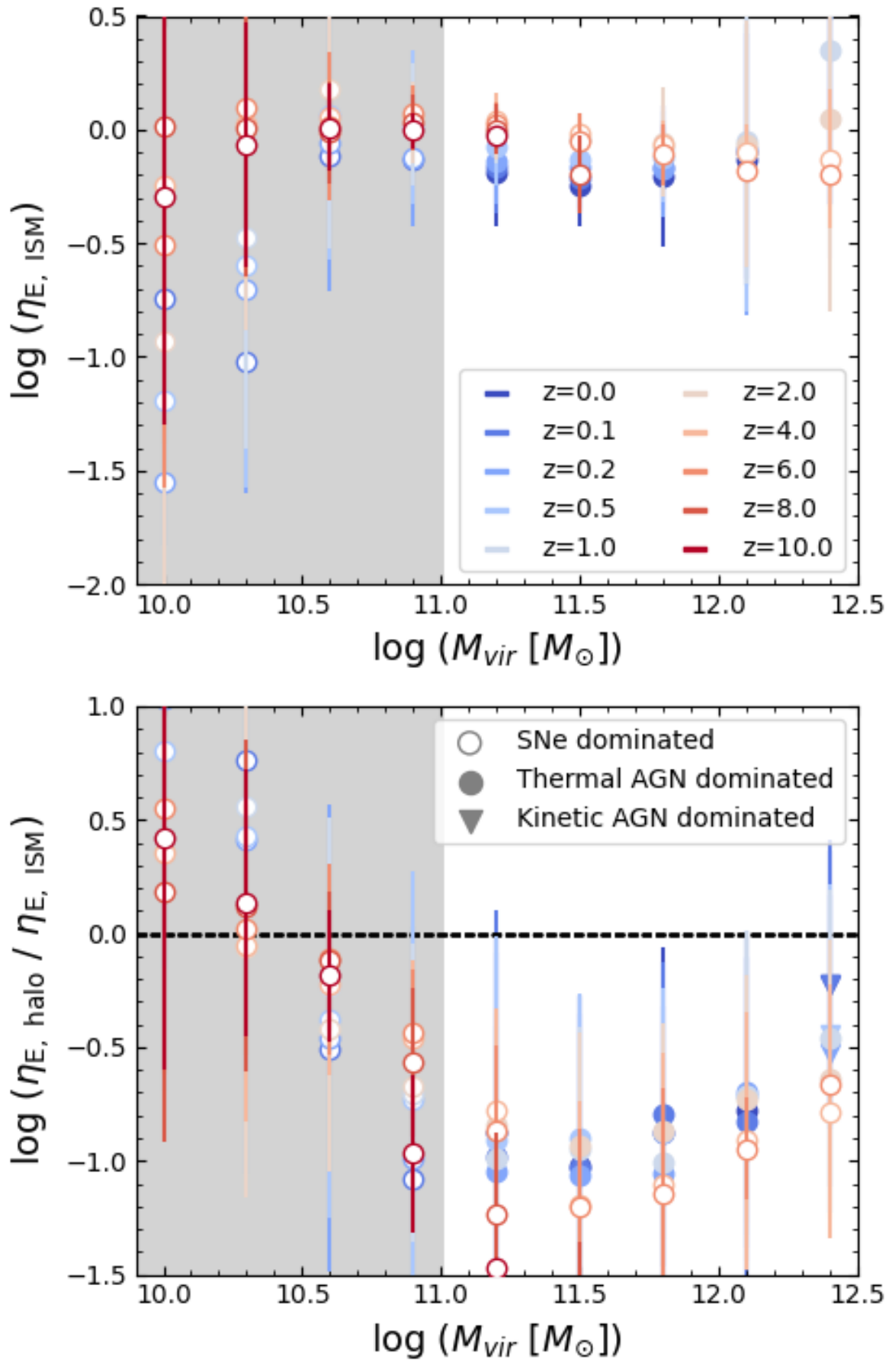

**Figure 13.** Top: energy loading factors at the ISM scale, shown for masses and redshift where kinetic AGN feedback is not the dominant feedback mechanism. $\eta_{E,\mathrm{ISM}}$ appears to vary less with the halo virial mass than $\eta_{M\ \mathrm{ISM}}$, with values ranging between ∼0.3 and ∼1 for most redshifts and halo virial masses. Bottom: ratio between the energy loading factors at the halo scale and the energy loading factors at the ISM shell, showing a very similar trend to the ratio between the mass loading factors at these scales. Colors and symbol types match those of Figure 5.

## 7. Metallicity

Lastly, we study the metal contents of the flows. In addition to discussing the metallicity of the gas flows themselves, we examine the relationship between the metallicity of the inflowing or outflowing gas and the reservoir from which it originates (the "metal enrichment factor"), and the ratio between the metallicity of the flow and the estimated metal yield from SNe (metal loading factor), all of which quantify the effectiveness with which gas flows carry and mix metals between the galaxy and the CGM and IGM.

### 7.1. Metallicity of Gas Flows

In Figure 14 we plot the flow metallicities, $Z \equiv \dot{M}_Z/\dot{M}$, where the metal mass flow rates $\dot{M}_Z$ are estimated using Equation (12), and the total gas mass flow rates $\dot{M}$ are estimated using Equation (9). We normalize metallicities relative to the solar metal mass fraction $Z_\odot = 0.0127$ (adopted from M. Asplund et al. 2009). The panel ordering, marker types, and color coding are as in Figure 5.

A surprising result, most likely due to resolution issues in low-mass TNG100 galaxies, is the dichotomy between halos with virial masses $\lesssim 10^{11}\,M_\odot$ and those above this mass, independent of shell or direction. At the low-mass end the metallicities increase with redshift (i.e., at $z = 10$ the metallicities are higher than in $z = 0$), but as we reach one of the mass bins near $10^{11}\,M_\odot$ the variation with redshift decreases and then flips, such that at higher masses, the metallicities decrease with redshift. Another expression of the same phenomenon is the dependence of the flow metallicities on the halo virial mass for a given redshift. For $z = 10$, the flow metallicity through a given shell and direction does not vary significantly with the halo virial mass, but as the redshift decreases the flow metallicity increases more sharply with the halo virial mass. This increase continues until the onset of kinetic AGN feedback is reached, after which the mass dependence of the flow metallicity either flattens or drops, depending on the direction of the flow and on the scale in which we measure it. At redshifts where kinetic AGN feedback is not dominant, the flow metallicities keep increasing with the halo virial mass.

For most virial masses and redshifts, ISM outflows have the highest metallicities, as they carry the metals formed in the galaxy. Except for dwarfs, ISM outflows at all times have higher metallicities than ISM inflows by roughly 0.2 dex. Outflows at the halo scale are the third highest, as they probably contain metals formed in the galaxy and carried by ejecta, mixed with lower-metallicity entrained CGM gas. Lastly, the metallicities of the inflows at the halo scale are the lowest relative to the others, as they are most likely to include only intergalactic, relatively pristine gas (though it is interesting that even at high redshift, gas inflowing into the CGM can be enriched up to more than $0.1\,Z_\odot$).

The dependence of the gas flow metallicities on the halo virial mass appears similar between all scales and directions, as the flow metallicities at both the ISM and halo scales increase with the halo virial mass until peaking at the onset of kinetic AGN feedback. The exception is the metallicities of the inflows at the halo scale, which remain relatively constant past the onset of kinetic AGN feedback. The metallicities of inflows and outflows at the ISM scale and of outflows at the halo scale all decrease (slightly) past the onset of kinetic AGN feedback. This is true for the reservoirs themselves as well, i.e., the ISM and CGM metallicities. The former result can also be seen in P. Torrey et al. (2019), who observed a drop of ISM metallicities in TNG100 galaxies more massive than $\sim 10^{10.5}\,M_\odot$, equivalent to the virial mass where we observe the onset of kinetic AGN feedback (see, e.g., Figure 7 of Y. Oren et al. 2024).

### 7.2. Metal Enrichment Factors

In addition to describing the metallicity of the flow, we wish to determine the relationship between the gas flow and the environment from which it originated. We therefore define a "metal enrichment factor"

$$\zeta_{\mathrm{flow}} \equiv \frac{\dot{M}_Z}{\dot{M} \cdot Z_{\mathrm{reservoir}}} = \frac{Z_{\mathrm{flow}}}{Z_{\mathrm{reservoir}}}, \qquad (16)$$

where $\dot{M}_Z$ is given by Equation (12) and $\dot{M}$ is given by Equation (9), such that $Z_{\mathrm{flow}}$ is the parameter we plot in Figure 14. $Z_{\mathrm{reservoir}}$ describes the average metallicity of the gas reservoir from which the outflows or inflows originate: the ISM metallicity for outflows at the ISM scale, and the CGM metallicity for outflows at the halo scale and for inflows at the ISM scale. Because the IGM metallicity is outside the scope of this work, we do not discuss $\zeta_{\rm in,halo}$. In most SAMs and gas regulator models, the metallicities of the reservoir and the flow are assumed to be identical, i.e., $\zeta = 1$ (e.g., K. Finlator & R. Davé 2008), while in others it is not (e.g., M. S. Peeples & F. Shankar 2011). As we mentioned in Section 3.1, the metal enrichment factor for SN-driven outflows in TNG when they are launched is a constant 0.4.[12]

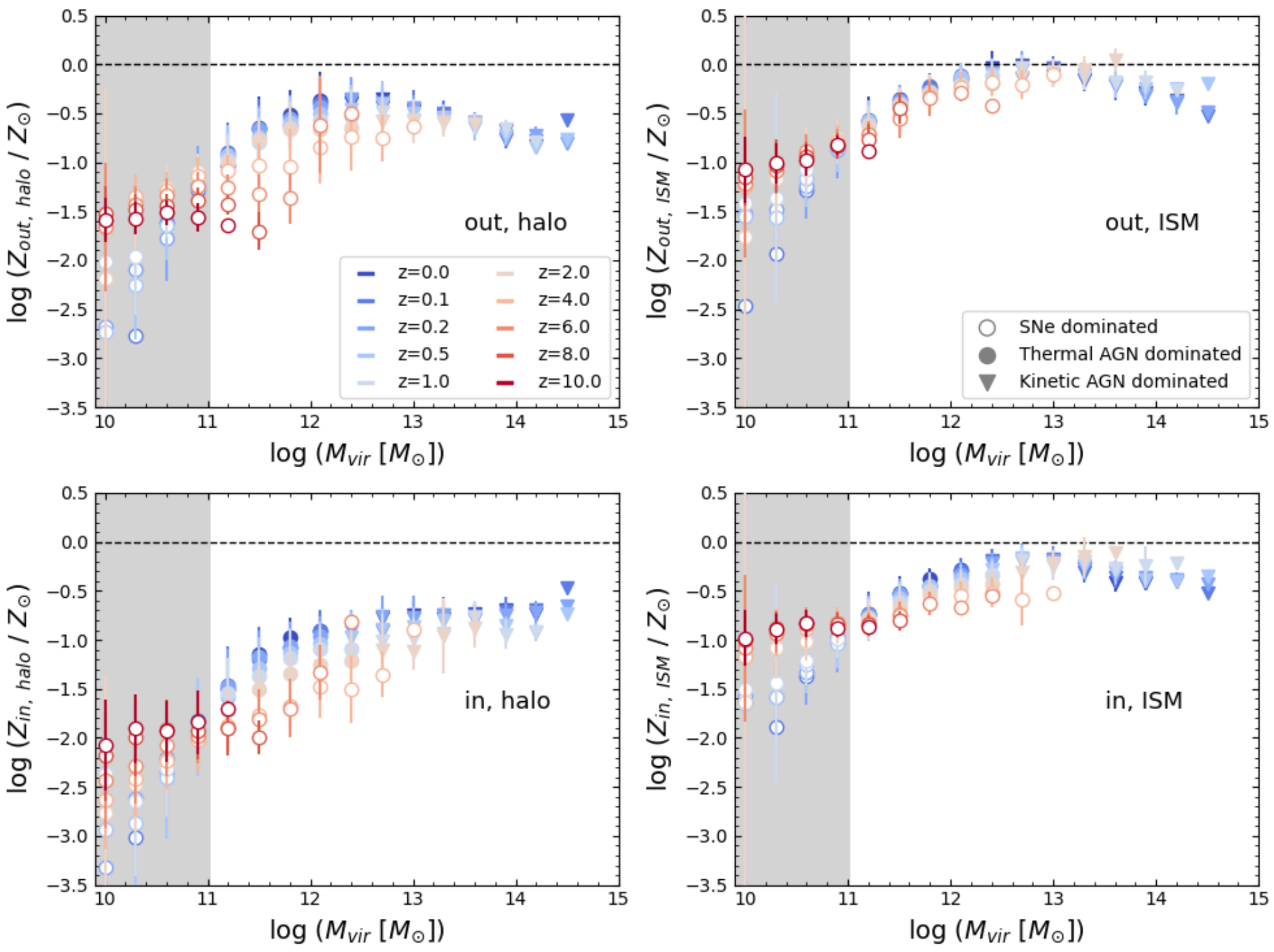

**Figure 14.** Metallicity of outflows (top row) and inflows (bottom row). The left column shows the halo scale, and the right column shows the ISM scale. Colors and symbol types match those of Figure 5. The metallicity of the flow rates increases with halo virial mass until the onset of kinetic AGN feedback, after which it either stabilizes with virial mass (galactic inflows, lower left panel) or decreases slightly (all other panels).

We start with the outflows at the ISM scale in the central panel of Figure 15, which shows a dichotomy between two dominant feedback mechanisms. In virial masses where outflows are driven by SNe, $M_{\rm vir} \lesssim 3 \times 10^{12}\, M_\odot$ (which includes halos dominated by thermal AGN feedback), $\zeta_{\rm out,ISM}$ ranges from 0.4 at $z = 0$ to 0.7 at $z = 0$, where the TNG metal enrichment factor at injection is 0.4. At high redshifts, where the CGM densities are also high, the dense surrounding gas may stall outflows and there is very little mixing of outflows with the surrounding gas, causing little to no change to the metal enrichment at injection. At low redshifts, however, we find evidence of entrainment (e.g., Section 5.4), which adds high-metallicity ISM to the winds and contributes to a higher enrichment factor. In virial masses where the outflows are driven by AGN, on the other hand, we find that $\zeta_{\rm out,ISM} = 1$ since kinetic AGN feedback simply gives additional momentum to ISM particles and does not alter their metallicity.

The metal enrichment factor for ISM inflows, shown in the right panel of Figure 15, suggests that ISM inflows have higher metallicities than the average metallicity of the CGM. For halo virial masses $\gtrsim 6 \times 10^{11}\, M_\odot$, there is an almost constant $\zeta_{\rm in,ISM} \approx 1.5$ for all redshifts. As the virial masses decrease, the metal enrichment factor increases slightly until reaching a peak of three for $M_{\rm vir} \approx 3 \times 10^{10}\, M_\odot$, and drops back for lower masses. To understand why the inflows have a seemingly higher metallicity than the medium through which they travel, we first consider that the metallicity of the CGM decreases with galactocentric radius (see, e.g., A. M. Garcia et al. 2023). In Y. Oren et al. (2024) we show that the density of the CGM at $z = 0$ can be very well approximated by a power law of the radius for halos with virial masses higher than $10^{11}\, M_\odot$. We can also fit the metallicity of the CGM to a power law of the radius, with a different power-law index than that of the density. If we use our power laws, we can estimate the ratio between $Z(0.1R_{\rm vir})$ and the average CGM metallicity between 0.1 and $1R_{\rm vir}$ ($\langle Z_{\rm CGM}\rangle$) as

$$\frac{Z(0.1R_{\rm vir})}{\langle Z_{\rm CGM}\rangle} = 0.1^{-a_Z}\,\frac{3 - a_n - a_Z}{3 - a_n}\,\frac{1 - 0.1^{3-a_n}}{1 - 0.1^{3-a_n-a_Z}},$$

where $a_n$ is the hydrogen density power-law index and $a_Z$ is the metallicity power-law index. For the median $z = 0$ halos in the $\log(M_{\rm vir}/M_\odot) = [11, 12, 13, 14]$ mass bins, we find $a_n = [1.6, 1.7, 1.2, 1.9]$ and $a_Z = [1.1, 0.8, 0.3, 0.5]$, giving the ratios [5.2, 3.4, 1.7, 2.1]—all very close to the numerically calculated $\zeta_{\rm in,ISM}$ for $z = 0$, supporting our conclusion that the metallicity of ISM inflows matches the metallicity of the CGM at the border between the two.

[12] In the works that describe the Illustris and the IllustrisTNG methods, such as M. Vogelsberger et al. (2013) and A. Pillepich et al. (2018a), what we call the "metal enrichment factor" is referred to as a "metal loading factor."

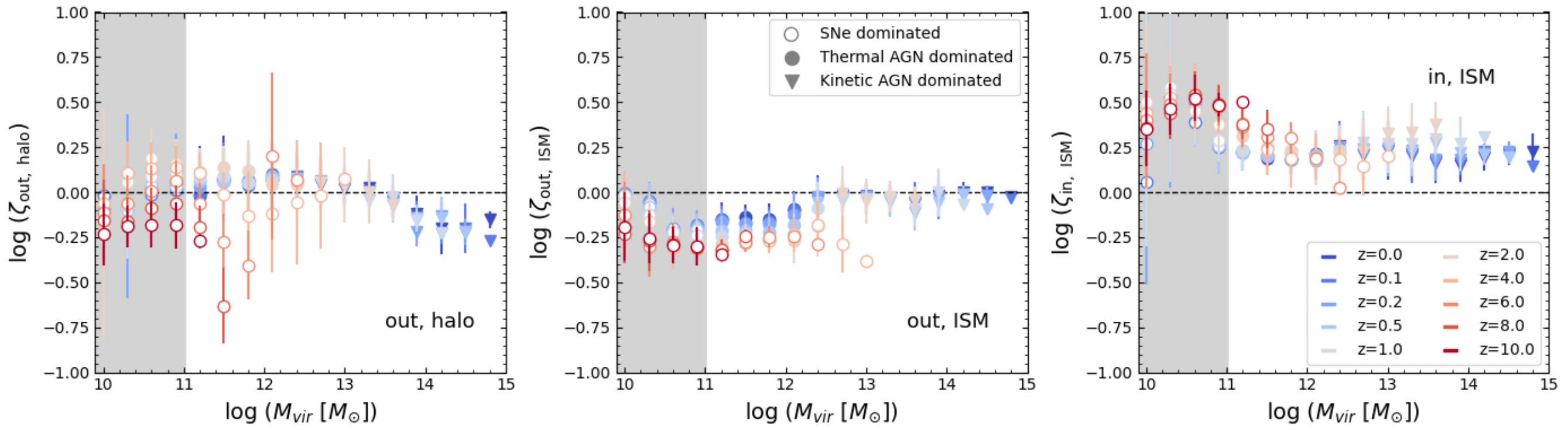

**Figure 15.** Metal enrichment, as defined in Equation (16). Colors and symbol types match those of Figure 5. Left panel: outflows at the halo scale. Generally, outflow metallicity matches the CGM metallicity, except for low-$z$, high-mass halos where the outflows have ∼50% fewer metals than the CGM average. Central panel: outflows at the ISM scale. SN-driven outflows appear to have a constant $\zeta = 0.6$ (slightly higher than the TNG metal loading at injection of 0.4), while AGN-driven outflows exactly match the metallicity of the ISM. Right panel: inflows into the ISM. The metallicity of the inflows appears to be 1.4–3 times higher than the average CGM metallicity, where the highest ratio is at the low-mass end.

Last is the metal enrichment of outflows at the halo scale, shown in the left panel of Figure 15. $\zeta_{\rm out,halo}$ suggests that the metallicities of the outflows from the halo are within a factor of 2 from the average metallicity of the CGM from which they emerge, barring two outliers (one at $z = 6$ and the other at $z = 8$). At low redshifts ($z \leqslant 2$), $\zeta_{\rm out,halo}$ appears to start at ∼0.75, then increases with the halo virial mass to ∼1.2, decreases toward the high-mass end, and flattens at ∼0.6 for $M_{\rm vir} \gtrsim 10^{14}\,M_\odot$. For $z \leqslant 1$ the decrease in metal enrichment starts after the onset of kinetic AGN feedback, and for $z = 2$, $\zeta_{\rm out,halo}$ starts to decrease at a lower mass, where the outflows are dominated by SNe. At high redshifts, however, there is no apparent trend with halo mass and $\zeta_{\rm out,halo}$ is either constant for all halo virial masses (as is the case for $z = 10$), or fluctuates around one.

At the highest-mass end there are net inflow rates of both mass (Figure 6) and energy (Figure 12) plus weakening preventative feedback (Figure 8), suggesting that much of the outflowing material mixes with the CGM and does not make it out of the halo. We would therefore expect that for halos more massive than $\sim 10^{14}\,M_\odot$ the metallicity of outflows at the halo scale will be similar to that of the CGM at the halo scale, and $\zeta_{\rm out,halo} = Z(R_{\rm vir})/\langle Z_{\rm CGM}\rangle$. Using the same power-law distributions described above, we get $Z(R_{\rm vir})/\langle Z_{\rm CGM}\rangle = 0.67$, in agreement with $\zeta_{\rm out,halo}$ from the simulation. While we cannot account for the metal enrichment factors at masses lower than those of galaxy clusters, we still find them to be relatively close to unity. Comparatively, the metallicities of the flows themselves can change by roughly 2 dex in a given redshift. We therefore conclude that the metallicity of halo outflows tracks the metallicity of the CGM to a good approximation.

### 7.3. Metal Loading

We finally describe the efficiency with which metals are ejected from the galaxy relative to a reference yield times the SFR. We observe in Figure 16 the metal loading factors at the ISM scale, estimated using Equation (15). Colors and markers follow our standard conventions.

For a given redshift, the metal loading factors of SN-driven outflows do not vary significantly with the halo virial mass above $\sim 3 \times 10^{10}\,M_\odot$. The metal loading factors out of galaxies in the centers of halos with virial masses smaller than $\sim 3 \times 10^{10}$ decrease with halo mass, with a slope that decreases as redshift increases, such that at the low-mass end, $\eta_Z$ for $z = 10$ is comparable to $\eta_Z$ for $z = 0$. For virial masses higher than the stated cutoff mass, the metal loading factor only increases with decreasing redshift.

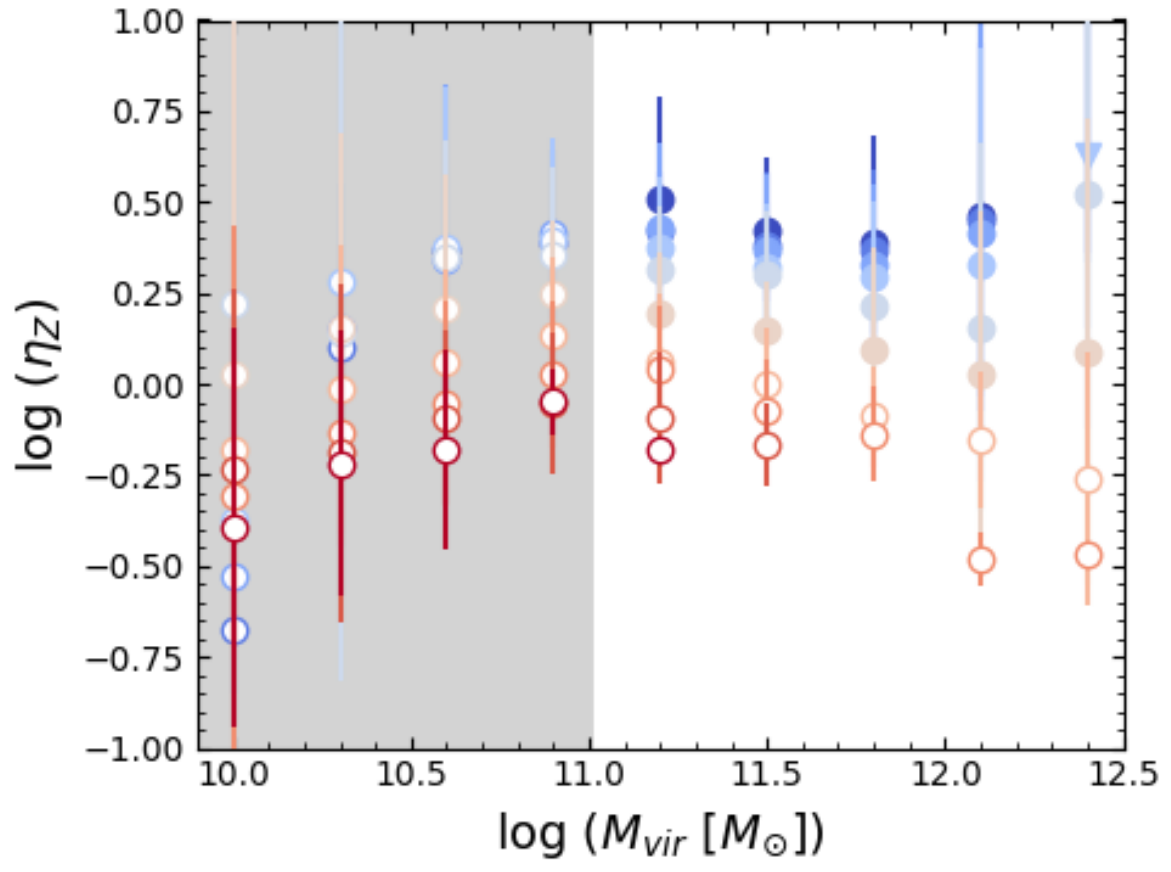

**Figure 16.** The metal loading factor based on Equation (15), estimated at the ISM scale. Colors and symbol types match those of Figure 5. For a given redshift, and for $\log(M_{\rm vir}/M_\odot) > 10.5$, $\eta_Z$ is roughly constant with virial mass. For a given virial mass, the metal loading factor decreases with increasing redshift.

The metal loading factor is larger than unity in the region where it flattens with halo virial mass, and for $z \leqslant 4$. It reaches ∼3 for $z = 0$, while for $z = 10$ it is ∼0.7. This would not necessarily suggest that at late times outflows carry more metals than could have been produced by stars, as $\zeta_{\rm out,ISM} \leqslant 1$ for all masses and redshifts, and is entirely calculated using the simulation results (whereas for $\eta_Z$ we assume a constant metal yield independent of halo mass or redshift). However, this indicates that for most halo masses, the outflows carry more metals out of the galaxy (per unit SFR) as time progresses. At the low-mass end $\eta_Z$ suggests that outflows carry more metals out of galaxies at earlier times.

## 8. Discussion

### 8.1. Baryon Flows at Different Halo Mass Scales

In our study we find distinct flow behaviors in halos of different total mass, due to the interplay between various feedback mechanisms and the gravitational potential. Within the range of halo masses that are well resolved and are represented in sufficient numbers in the TNG100 volume, we can identify three main regimes: low-mass halos ($10^{11} \lesssim M_{\rm vir} \lesssim 10^{12.25}\,M_\odot$), intermediate-mass halos ($10^{12.25} \lesssim M_{\rm vir} \lesssim 10^{13.5}\,M_\odot$), and massive halos ($M_{\rm vir} \gtrsim 10^{14}\,M_\odot$).

#### 8.1.1. Low-mass Halos

Low-mass halos ($10^{11} \lesssim M_{\rm vir} \lesssim 10^{12.25}\,M_\odot$) are inhabited mostly by galaxies that are star forming at all redshifts (Figure 2), with outflows that are dominated by stellar feedback. Although most of these galaxies contain BHs, they are mostly in the thermal AGN feedback regime (see Figure 4). The gas depletion times are less than the Hubble time at all redshifts, leading to efficient star formation (Figure 9). Low-mass halos show net inflows of both mass and energy on halo scales at all redshifts, while at the ISM scale they gain mass but simultaneously lose energy. The net energy loss is likely due to cooling of the inflows before they reach the ISM scale. Baryon fractions are strongly subuniversal (Figure 7), show a weak trend with halo mass over this interval, and have decreased by almost a factor of 2 over cosmic time since $z \sim 10$. At halo scales, low-mass halos do not show signs of preventative feedback at high redshift, and show only weak preventative feedback at $z = 0$ (Figure 8). On ISM scales, low-mass halos show signatures of "galactic fountains," as inflows into the ISM shells surpass those into the halo shells, suggesting previously ejected material is able to fall back onto the galaxy in addition to fresh gas arriving from the IGM. This phenomenon is considered a crucial source of mass inflow into star-forming galaxies, as observed in our own Milky Way (J. N. Bregman 1980) and in nearby galactic halos (e.g., F. Fraternali & J. J. Binney 2006, 2008). Figure 8 suggests that for $z \geqslant 4$ there is an extra source of inflow at the halo scale, which we interpret as a "halo fountain."

A very significant factor in the occurrence of a galactic fountain is the ability of the ejecta to cool in order to fall back onto the galaxy. The virial temperature associated with the halos in this mass range, which is also the typical temperature of the volume filling phase of the CGM, is in the region of $\sim 10^5$ K (see, e.g., Y. Oren et al. 2024). This temperature is associated with the peak of the cooling curve (e.g., O. Gnat & A. Sternberg 2007; R. P. C. Wiersma et al. 2009), and therefore CGM gas at this temperature can cool very effectively. Additionally, this may explain why we do not witness a galactic fountain in higher-mass halos—higher virial temperatures lead to slower cooling, such that outflows are either ejected entirely from the halo or simply mix with the thermally supported hot CGM.

#### 8.1.2. Intermediate-mass Halos

We define intermediate-mass halos as those with halo virial masses between $10^{12.25}$–$10^{13.5}\,M_\odot$, where the onset of both kinetic BH feedback (Figure 4) and galaxy quenching (Figure 2) occurs. Before the onset of kinetic AGN feedback, we measure net positive inflow rates on halo scales in both mass and energy (Figure 12). The ISM depletion times are very short, as low as $\sim$10% of the Hubble time, leading to the rapid formation of stars and the buildup of metals.

After the onset of kinetic AGN feedback, at $z \leqslant 2$, the picture changes dramatically. The mass and energy flows become net positive outward at both the ISM and halo scales. The baryon fractions of both the CGM and the entire halo decrease, supporting a picture of AGN driving outflows beyond the virial radius. The SFRs drop, and the ISM depletion times increase. As this occurs, the metallicities begin to drop as fewer new stars are formed and outflows carry the existing metals out of the galaxy. E. Zinger et al. (2020) have discussed how the onset of kinetic AGN feedback in TNG100 galaxies at $z = 0$ is associated with rapid quenching of star formation, whereas P. Torrey et al. (2019) also attribute the drop in metallicity of the ISM to the "increased impact of AGN feedback," and M. S. Peeples & F. Shankar (2011) invoke outflows for their model to reproduce the observed drop in the mass–metallicity relation past Milky Way–like galaxies. The metal enrichment factor of ISM outflows transitions from $\sim$0.4, which is the metal enrichment factor of SN outflows at injection, to $\sim$1, indicating the outflows now predominantly contain ISM particles kicked by kinetic AGN feedback and not wind particles driven by SNe.

We see the strongest preventative feedback on halo scales in this halo mass range, with the gas inflow rate suppressed relative to $f_{\rm B}\dot{M}_{\rm in}^{\rm DM}$ by about 0.25 dex. The gas inflow rate at ISM scales drops dramatically relative to the DM halo growth rate (Figure 8). The ISM gas inflow rate $\dot{M}_{\rm in,ISM}$ also begins to drop below the halo-scale gas inflow rate $\dot{M}_{\rm in,halo}$, indicating the buildup of a hot CGM.

#### 8.1.3. High-mass Halos

High-mass halos include all halos with virial masses higher than $10^{13.5}\,M_\odot$. These halos are all dominated by kinetic AGN feedback, and at low redshift they are mostly occupied by quenched galaxies. Note that there are very few halos in this mass range at high redshift in the TNG100 volume (Figure 1), so most of the trends discussed below are measured for redshifts $z \lesssim 1$. There is a strong net inflow of mass into high-mass halos on halo scales, while on ISM scales the mass inflow and outflow rates are close to equal. The baryon fractions increase strongly with increasing halo mass, from about half to 80% of the universal value by $M_{\rm vir} \simeq 10^{14}$–$10^{15}\,M_\odot$. Halo-scale preventative feedback is still present, but it is somewhat weaker than in intermediate-mass halos, and continues to weaken with increasing halo mass. ISM-scale preventative feedback is strongest for halos with mass $= 3 \times 10^{13}\,M_\odot$, and from there it weakens as halos and galaxies become more massive. In the highest-mass bin we find that $\dot{M}_{\rm in,halo} \simeq \dot{M}_{\rm in,ISM} \simeq f_{\rm B}\dot{M}_{\rm in,DM}$, suggesting that AGN feedback is beginning to fail to be able to regulate accretion. High-mass halos show a net inflow of energy at the halo scale, while at the ISM scale there is net energy loss from $M_{\rm vir} \sim 10^{13}$–$10^{14}\,M_\odot$, while $\dot{E}_{\rm in} \simeq \dot{E}_{\rm out}$ above $10^{14}\,M_\odot$, with a trend toward net energy inflows toward the highest halo masses. Outflows in these high-mass halos are also depleted of metals relative to the overall average metallicity of the CGM (left panel of Figure 15).

### 8.2. Interactions between Outflows and the Surrounding Gas—Entrainment, Mixing, and Stalling

The way in which injected momentum or energy couples with the medium through which winds propagate has a large impact on the emergent properties of the outflows. Comparing the mass loading factor, $\eta_M$, at the ISM shell to the mass loading factor at injection, $\eta_{w,\,\rm TNG}$, based on the TNG subgrid recipe (left panel of Figure 10), shows that the ratio between the two decreases with redshift, such that for a given halo mass, the emergent mass flow rates are weaker relative to the amount of injected wind material at earlier times. This result is independent of the halo virial mass, leading us to suggest that this effect is due to higher gas densities in the early universe, where the gas density of the CGM is proportional to $\rho_c \propto (1+z)^3$. The higher densities at high redshift could cause winds to stall, lowering $\eta_M$. At low redshift, the winds may entrain additional material, increasing $\eta_M$. Evidence of significant amounts of CGM gas entrained by galactic outflows at low $z$ were found in the EAGLE simulation, in agreement with our interpretation, where P. D. Mitchell et al. (2020b) show that at $0.1R_{\rm vir}$, 50% of outflowing particles are entrained CGM gas.

Additional evidence suggesting that winds stall at higher redshifts (leading to the deceleration of outflows) and entrain CGM material at lower redshifts comes from our analysis of the metal enrichment factors of SN-driven outflows at the ISM shell. The central panel of Figure 15 shows that the metal enrichment factor increases from 0.4 at $z=10$—the metal enrichment factor at injection, indicating no mixing with the surrounding CGM—to 0.7 at $z=0$, suggesting SN outflows carry surrounding material with them. This enrichment value is the average between 0.4 ($\zeta$ of wind particles only) and one ($\zeta$ of ISM particles only), in agreement with the EAGLE result that 50% of the outflows at the ISM scale consists of entrained gas.

### *8.3. Comparison with Previous Works*

In this section, we briefly discuss how our results compare and contrast with select results from the literature, including gas regulator models, SAMs, and numerical hydrodynamic simulations.

#### *8.3.1. Gas Regulator Models*

Gas regulator models track flow quantities between different reservoirs (such as the CGM, ISM, and stars), and generally include the growth of the main progenitor of a DM halo based on *N*-body simulations or fitting functions derived from them. They are similar to SAMs, but in general are more empirical and do not include the full DM merger history (including nonmain branches) or the physical processes associated with galaxy mergers.

Most gas regulator models in the literature track only the flows of mass and metals, and do not track energy flows. C. Carr et al. (2023) were the first to include energy flows in a gas regulator model, and showed that energy flows can play a central role in regulating star formation via regulating cooling of the CGM. In their fiducial model, which was adjusted to approximately reproduce the stellar mass to halo mass (SMHM) relation from P. Behroozi et al. (2019), they adopted mass and energy loading factors that are power-law functions of the halo mass. The values of the mass loading range from $\eta_M = 10$ for $10^{10}\,M_\odot$ halos to $\eta_M = 1$ for the $10^{12}\,M_\odot$ halos, and the energy loading from $\eta_{\rm E} \sim 1$ for the low-mass end to $\sim$0.1 for the high-mass end. The adopted mass loading factors in C. Carr et al. (2023) are much lower than the values that we measure in TNG, and the dependence of energy loading on halo mass is also very different ($\eta_{\rm E}$ has only a weak dependence on halo mass in TNG).

C. Carr et al. (2023) also implement a halo-scale preventative feedback parameter, $f_{\rm prevent}$, used to moderate baryonic inflows into the halo relative to DM inflows. As the energy flows in C. Carr et al. (2023) are purely thermal, $f_{\rm prevent}$ is also used to regulated energy inflows into the halo relative to energy outflows. The $f_{\rm prevent}$ obtained from the calibration to the P. Behroozi et al. (2019) SMHM relation ranges from 0.3 at the low-mass end to 0.8 at the high-mass end, so while it is by definition capped at one this limit is not reached. In comparison, the left panel of Figure 8 suggests a somewhat different picture in TNG100. $\dot{M}^{\rm gas}_{\rm in,halo}/f_{\rm B}\dot{M}^{\rm DM}_{\rm in,halo}$ is close to unity or even slightly above it (indicating no halo-scale preventative feedback or halo fountains) at $z \geqslant 2$, and at $z=0$ it has a nearly constant value of $\simeq$0.7 at the lowest well-resolved halo up to about $M_{\rm vir} \sim 10^{12}\,M_\odot$, where it begins to decline due to the onset of kinetic AGN feedback.

Another gas regulator model we wish to discuss here was published in a set of companion papers, G. M. Voit et al. (2024a, 2024b). Unlike previous regulator models, the Voit model tracks all mass and energy associated with a halo's baryons, even those that may have been pushed beyond the halo's virial radius by feedback, and the energy budget manifestly includes the halo's gravitational potential energy. A key quantity in regulating gas accretion and hence star formation is the specific energy of the outflow $\varepsilon_{\rm fb} \equiv \eta_{\rm E}\varepsilon_{\rm SN}/\eta_M$, where $\varepsilon_{\rm SN}$ is a constant specific energy attributed to SN feedback.[13] They then distinguish between coupled and uncoupled outflows. A coupled outflow can transfer energy to the CGM causing preventative feedback, while uncoupled outflows cannot and therefore cause only ejective feedback. The key idea of the Voit model is that for coupled outflows, the ratio between $\varepsilon_{\rm fb}$ and the specific energy of the accreting gas $\varepsilon_{\rm acc}$ determines the future of the galaxy. Namely, if $\varepsilon_{\rm fb} > \varepsilon_{\rm acc}$ then the halo will expand (leading to reduced cooling and star formation), and if $\varepsilon_{\rm fb} < \varepsilon_{\rm acc}$ it will contract (leading to enhanced cooling). If $\varepsilon_{\rm fb}$ drops below $\varepsilon_{\rm acc}$, a contraction crisis occurs where the halo keeps contracting and runaway star formation may occur. To prevent this, the halo will require a new source of feedback to counter the collapse—such as AGN feedback.

G. M. Voit et al. (2024a) discuss the interpretation of IllustrisTNG in the context of their model (see especially their Figure 3) and suggest that for $z \gtrsim 4$, $\varepsilon_{\rm fb} < \varepsilon_{\rm acc}$ and so halos should be in the contracting regime and therefore strongly cooling and star forming. This is qualitatively consistent with our finding that in the mass range considered in G. M. Voit et al. (2024a)'s Figure 3 ($5\times10^{11}$–$7\times10^{12}\,M_\odot$), there are strong net inflows into halos at $z \gtrsim 4$, and these weaken over time as $\epsilon_{\rm fb}/V^2_{\rm vir}$ increases (due to the assumed scaling in the IllustrisTNG stellar wind subgrid model), pushing halos in the more strongly "expanding" regime. However, we note that G. M. Voit et al. (2024a) assume that the emergent wind energies are the same as the wind launch properties, while we have shown that emergent wind properties can be quite different due to entrainment and stalling, and these differences can change with the conditions in the CGM (and thus appear different at different cosmic times).

[13] Estimating $\varepsilon_{\rm fb}$ for SN-driven outflows in TNG100 using Equations (4), (13), and (14) gives a similar relationship, with $e_w$ taking the place of $\varepsilon_{\rm SN}$. Note that $e_w$ depends on metallicity while $\varepsilon_{\rm SN}$ is a constant.

#### 8.3.2. Semianalytic Models

Traditional SAMs solve ODEs describing flows of gas mass and metals between the IGM, CGM, and ISM, and the rate at which ISM gas is converted into stars (see R. S. Somerville & R. Davé 2015, for a review). As noted above, SAMs typically include more physically motivated model components than gas regulator models, and generally account for the full DM halo formation history, thus including the physical processes associated with galaxy mergers.

P. D. Mitchell et al. (2020b) show a useful comparison of the mass loading factors as a function of halo mass for several SAMs from the literature in their Figure 13, including GALFORM (P. D. Mitchell et al. 2018), L-galaxies (B. M. B. Henriques et al. 2015), the Santa Cruz SAM (R. S. Somerville et al. 2015), GAEA (M. Hirschmann et al. 2016), and SHARK (C. d. P. Lagos et al. 2018). Some SAMs do not allow flows directly from the CGM to the IGM (i.e., halo outflows); for those that do include halo outflows, both halo-scale and galaxy-scale mass loading is shown. It is also important to note that most traditional SAMs do not include halo-scale preventative feedback arising from internal stellar or AGN feedback, i.e., it is generally assumed that the mass inflow rate into the halo is $f_{\rm B}\dot{M}^{\rm DM}_{\rm in,halo}$.[14] Although all of these SAMs are calibrated to reproduce the $z\sim0$ stellar mass function and stellar mass versus halo mass relationship, this figure shows that there is an extremely wide range of slopes and normalizations for the mass loading of stellar-driven winds in different SAMs. We notice that SAMs that do not include halo outflows (e.g., GALFORM and the Santa Cruz SAM), in which star formation must be regulated solely by ejecting gas from the ISM, unsurprisingly must invoke very large mass loadings, which increase steeply toward lower halo masses. These mass loadings ($\eta\sim10$–$100$ at $M_{\rm vir}\sim10^{11}\,M_\odot$) are a bit higher than the ones we measure in TNG100 over the same halo mass range. The models that include halo-scale outflows (L-galaxies and SHARK) adopt somewhat lower ISM-scale mass loadings. The GAEA SAM was one of the first SAMs to include halo-scale preventative feedback, and it adopts the lowest mass loadings—even lower than either our TNG100 measurements or those from EAGLE (presented in Figure 13 of P. D. Mitchell et al. 2020b).

V. Pandya et al. (2020) presented a detailed comparison of halo- and galaxy-scale inflow and outflow rates, along with ISM, stellar, and CGM masses between the Santa Cruz SAM and the FIRE-2 simulations (P. F. Hopkins et al. 2018). They showed that the mass inflow and outflow rates in the Santa Cruz SAM were much larger than those predicted by FIRE, especially in low-mass halos. This lead to the strong conclusion that SAMs may need to adopt more flexible assumptions about galaxy- and halo-scale inflows and outflows (including the option of halo-scale preventative feedback and halo-scale outflows) in order to faithfully reproduce the baryon cycle in numerical simulations.

This discussion highlights that there are many different baryon cycle configurations that can lead to the same stellar mass versus halo mass relation in galaxies. Furthermore, several different aspects of the baryon cycle beyond wind mass loading out of the galaxy are also important: the mass outflow rate at the halo scale, the recycling time for mass that has been ejected from the halo, and preventative feedback on both the halo and galaxy scales. As discussed above, energy flows (neglected in traditional SAMs) can also play a crucial role in the baryon cycle. Metal flows (generally assumed to trace mass flows in a simple way in traditional SAMs) can also play a role by affecting cooling rates.

V. Pandya et al. (2023) presented the first SAM that was designed to reproduce the detailed baryon flow cycle measured from a high-resolution hydrodynamic simulation. Instead of adopting relatively arbitrary choices for many aspects of the baryon cycle, and then calibrating the free parameters to match a limited set of quasi observables at $z=0$ as in nearly all previous SAMs, V. Pandya et al. (2023) used measurements of the mass, metal, and energy inflow and outflow rates on halo (CGM) and ISM scales from the FIRE-2 zoom-in simulations to calibrate their SAM. The model also accounted separately for heating and turbulence driven by SN winds and cosmic accretion (i.e., thermal and kinetic energy). As in the simpler regulator models of C. Carr et al. (2023) and G. M. Voit et al. (2024b), they found that the specific energy of the SN-driven winds plays a critical role in regulating star formation by overpressurizing the CGM, leading to halo outflows and halo-scale preventative feedback, and by slowing down cooling from the CGM to the ISM (galaxy-scale preventative feedback). Turbulence can also play a significant role in preventing cooling at early times. Importantly, V. Pandya et al. (2023) showed that, when designed and calibrated in this way, the simplified framework of an SAM could accurately reproduce the main state variables (e.g., stellar mass, ISM mass, and CGM mass) and their derivatives (inflow and outflow rates at the galaxy and halo scales) from a numerical hydrodynamic simulation, motivating this work and that of O. Omoruyi et al. (2026, in preparation).

#### 8.3.3. Zoom-in Simulation: FIRE-2

We start our comparison with other simulations with FIRE-2, analyzed by V. Pandya et al. (2021). FIRE-2 is a zoom-in simulation, in which a region inside a low-resolution cosmological simulation is selected and evolved again in a much higher resolution, while the boundary conditions are kept from the original simulation.[15] The implementation of SN feedback in FIRE-2 is very different than that of TNG100: in FIRE-2, the probability of an SN explosion is estimated per star-forming particle and time step, according to which a feedback event may then occur. The SN event deposits energy, mass, and metals directly into the surrounding gas particles, and the fraction of thermal to kinetic energy is calculated per neighboring gas particle as well. Additionally, FIRE-2 does not include SMBHs or AGN feedback (but see D. Anglés-Alcázar et al. 2017b; P. F. Hopkins et al. 2023; S. Wellons et al. 2023). Due to the latter point we will limit our comparison to TNG100 halos in the $10^{10}$–$10^{12}\,M_\odot$ region.

V. Pandya et al. (2021) computed similar mass, energy, and metal loading factors to those we estimate in this work. Unlike this work, which defines outflowing particles as having $v_r>0$, V. Pandya et al. (2021) define outflows as particles with $v_B>v_{\rm esc}$, where $v_{\rm esc}$ is the escape velocity at the ISM shell. Under this definition fewer particles are considered as outflows, which effectively lowers $\eta_M$ and $\eta_Z$ by a factor of 2 relative to what they were using our definition.[16] Second, the loading factors are presented as a function of either stellar mass or virial velocity rather than of halo virial mass. We convert the stellar mass to a halo mass using the median ratio presented in P. F. Hopkins et al. (2018). In FIRE-2, the equivalent stellar mass range to our analysis is $10^7$–$10^{10.7}\,M_\odot$.

[14] Many SAMs do include a form of halo-scale preventative feedback arising from heating by the metagalactic UV background after reionization, but this tends to impact only very low-mass halos of $M_{\rm vir}\lesssim10^{10}\,M_\odot$.

[15] FIRE-2 gas particles can reach masses of $\sim$250–7100 $M_\odot$, compared to the TNG100 mass resolution of $\sim10^6\,M_\odot$ per gas cell.

The mass loading factor out of FIRE-2 galaxies reaches $\sim$20 at the low-mass end ($\sim$5 times lower than for TNG100), whereas for the high-mass end it drops below the TNG100 results and is closer to $\sim$0.1 at $z = 0$ (for TNG100, it is $\sim$10). Interestingly, the FIRE-2 mass loading factor for a given stellar mass does not appear to vary significantly with redshift, while in TNG100 we find a clear trend of increasing $\eta_M$ as the redshift drops. If we reorganize the loading factors as a function of the virial velocity rather than of the mass we find that for a given $V_{\rm vir}$ bin the redshift dependence of TNG100 mass loading factors significantly decreases. In FIRE-2, however, $\eta_M$ increases with redshift for a given $V_{\rm vir}$ bin, but the larger $V_{\rm vir}$ is, the larger the dynamic range covered by $\eta_M$. This is because for a given mass bin $V_{\rm vir}$ increases with redshift,[17] and thus the loading factors from a given mass bin will "skew to the right" as higher virial velocities are attributed to the same mass at higher redshifts.

Energy loading factors in FIRE-2 vary in their dependence on galaxy mass in different redshift bins, but qualitatively they decrease with galactic mass and increase with redshift. For a given mass bin, $\eta_{\rm E}$ in FIRE-2 may increase by roughly 1 dex from $z = 0$ to $z = 4$, and more specifically range from $\sim$0.1 to $\sim$1 at the low-mass end, to $\sim$0.02 to $\sim$0.2 at the high-mass end. As a function of the virial velocity, however, the FIRE-2 energy loading factors appear to be constant for $V_{\rm vir} \lesssim 100\,{\rm km\,s^{-1}}$ in both the intermediate- and high-redshift bins, after which they decrease, while the low-redshift $\eta_M$ appears to increase with $V_{\rm vir}$. Comparatively, the TNG100 energy loading factors are roughly 0.5 for most masses and redshifts (except for low masses, where they also increase with redshift), and therefore behave similarly as a function of the virial velocity.

The metal loading factors from the FIRE-2 simulation are quite different compared to those in TNG100. In FIRE-2, the metal loading factors remain at a relatively constant $\sim$1 until reaching $M_* \approx 10^9\,M_\odot$, after which they decrease until reaching $\sim$0.05 for $M_* \approx 10^{11}\,M_\odot$. As with the mass loading factors, there is no significant dependence on redshift for a given mass bin. Since $\eta_M$ is relatively constant for a wide range of stellar masses, as we reorganize it as a function of the virial velocity we find it to have a similar behavior—constant for $V_{\rm vir} \lesssim 100\,{\rm km\,s^{-1}}$ and decreasing afterward, with no significant variation with redshift. The metal loading factor of TNG100 galaxies, however, remains constant with halo mass (for $M_{\rm vir} \gtrsim 3 \times 10^{10}\,M_\odot$) until reaching the onset of kinetic AGN feedback, while for a given mass bin it may vary by 1 dex for different redshifts, going from $\sim$3 at $z = 0$ down to $\sim$0.5 at $z = 10$.

The metallicities of gas flowing into TNG100 halos are very close to those of FIRE-2, but only in the $100\,{\rm km\,s^{-1}}$ virial velocity bin.[18] In FIRE-2, $Z_{\rm in,halo}$ keeps decreasing with the virial velocity at a similar pace for all redshifts, while for TNG100, we find that low-$z$ metallicities drop much faster with $V_{\rm vir}$ than high-$z$ metallicities, as is the case in Figure 14. The preventative feedback parameter estimated for FIRE-2 is also very different from its TNG100 equivalent: on one hand, the trend with redshift seen in TNG100 almost vanishes as we consider $f_{\rm prev}$ as a function of the virial velocity, which is also the case for FIRE-2; on the other hand, this parameter never rises above one in FIRE-2, while it does so in $z \geqslant 2$ in TNG100.

#### 8.3.4. Cosmological Simulations

D. Nelson et al. (2019b) analyzed the outflow rates out of galaxies in TNG50, a higher-resolution counterpart of TNG100. Gas particles in TNG50 are roughly 2 dex less massive than those of TNG100. To allow for a higher resolution the box side length was decreased by half, leading to fewer high-mass halos appearing in TNG50 compared to TNG100 (e.g., at $z = 0$, only 35 TNG50 halos have virial masses $> 10^{13}\,M_\odot$ compared to 232 in TN100). On the other hand, lower-mass halos are much more abundant and more clearly resolved, allowing them to be investigated in more detail. D. Nelson et al. (2019b) discuss the mass loading factors of $z = 2$ galaxies in TNG50, measured at a 10 kpc distance from the galaxy. They also present their results as a function the stellar mass within a 30 kpc aperture, which differs from our definition of stellar mass used so far (relative to the stellar half-mass radius). The SFRs used for the mass loading factors are also taken within the same 30 kpc radius.

In Figure 17 we compare the mass loading factors from D. Nelson et al. (2019b) to the TNG100 loading factors at the ISM scale from $z = 2$. For an apples-to-apples comparison we estimate the stellar mass and SFRs in a 30 kpc aperture as well.[19] We mark the stellar mass (measured within 30 kpc) of galaxies whose associated virial radius is approximately 100 kpc with a vertical gray dashed line, and the stellar mass whose associated virial mass is the onset of kinetic AGN feedback for $z = 2$ with a vertical black dashed line. The gray shaded region that marks possibly unresolved galaxies in TNG100 extends to $\sim 10^{8.6}\,M_\odot$ rather than the $10^8\,M_\odot$ mentioned in Section 2.1 since under the new definition the galaxy limits extend farther and therefore include more stellar particles. For the SFRs, however, there is no significant contribution from the region between twice the stellar half-mass radius and a 30 kpc aperture.

The TNG50 mass loading factors are larger by $\sim$0.2 dex than the TNG100 mass loading factors up to the onset of kinetic AGN feedback. We note that for a given stellar mass bin up to the onset of kinetic AGN feedback, the TNG50 SFRs are comparable to their TNG100 counterparts. This therefore suggests that the SN-driven mass outflow rates in TNG50 galaxies are higher relative to their TNG100 counterparts.

Next we compare with the EAGLE cosmological hydrodynamic simulation (J. Schaye et al. 2015), whose outflow rates have been analyzed in P. D. Mitchell et al. (2020b) and inflow rates have been analyzed in P. D. Mitchell et al. (2020a). The

16 A comparison between $\eta_{\rm E}$ for the two definitions of outflows was not made explicitly by V. Pandya et al. (2021).
17 For a given halo virial mass, $V_{\rm vir} \propto R_{\rm vir}^{-1/2} \propto (1 + z)^{1/2}$.

18 $V_{\rm vir} = 100\,{\rm km\,s^{-1}}$ is proportional to $M_{\rm vir} \approx 5 \times 10^{11}\,M_\odot$ at $z = 0$, $M_{\rm vir} \approx 6 \times 10^{10}\,M_\odot$ at $z = 4$, and $M_{\rm vir} \approx 2 \times 10^{10}$ at $z = 10$.
19 We use instantaneous SFRs whereas D. Nelson et al. (2019b) take an average over the last 100 Myr. We find the difference between the two to be negligible for TNG100 galaxies.

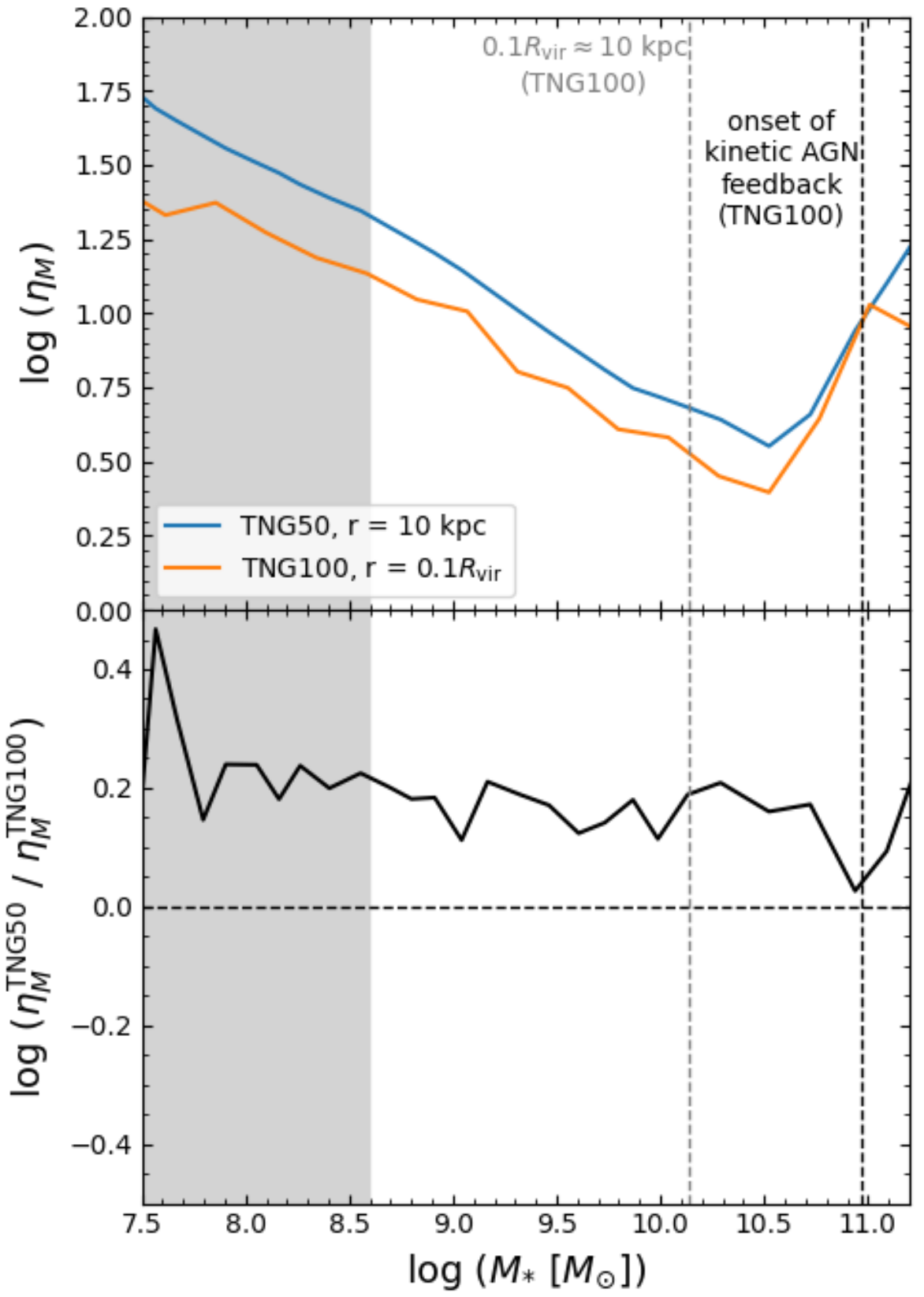

**Figure 17.** A comparison between the mass loading factors of TNG50 (blue line) and TNG100 (orange line) at $z = 2$, estimated as a function of the stellar mass within a 30 kpc aperture. The TNG50 mass loading factors, taken from D. Nelson et al. (2019b), are estimated at a distance of 10 kpc from the galactic center while the TNG100 mass loading factors are estimated at $0.1R_{vir}$. A gray dashed line marks the stellar mass associated with $R_{vir} \approx 100$ kpc (where $0.1R_{vir} = 10$ kpc). A black dashed line marks the stellar mass associated with the onset of kinetic AGN feedback at $z = 2$ in TNG100.

EAGLE simulation is very similar to Illustris-TNG100 in its method for solving hydrodynamics, its subgrid physics, its box size, and its resolution. The halos and galaxies that form within the EAGLE simulation span a similar mass range to those of TNG100. SN and AGN feedback, however, are implemented differently. SN feedback is purely thermal, and rather than launching decoupled wind particles, a star-forming gas particle adds thermal energy to its neighboring gas particles such that their temperature increases by $\Delta T = 10^{7.5}$ K. BH particles are seeded in galaxies in a similar fashion to those of IllustrisTNG (i.e., planted within halos with $M_{vir} > 10^{10} M_\odot$), and are allowed to accrete material at the Bondi accretion rate, capped by the Eddington accretion rate. SMBH feedback is implemented similarly to SN feedback, via injection of pure thermal feedback that is applied to neighboring particles. The energy injected into nearby particles is also proportional to $\dot{M}_{BH}c^2$, and is accumulated until it is large enough to heat a nearby gas particle by $\Delta T = 10^{8.5}$ K, after which a feedback event occurs.

The normalized mass flow rates are relatively close to those of TNG100, and the TNG100 values are higher by at most 1 dex for all redshifts and scales. More can be learned from the mass dependencies. At the ISM scale of EAGLE galaxies, both the inflow and outflow rates peak at $M_{vir} \approx 5 \times 10^{11} M_\odot$, similarly to the ISM inflow and outflow rates in TNG100 (in TNG100 this occurs at a slightly lower mass). In EAGLE, the peak in inflow rates is correlated with a relatively high recycling fraction at the ISM scale, similar to our identification of a galactic fountain in TNG100 at this mass scale (P. D. Mitchell et al. 2020a). However, SN-driven outflows in TNG100 do not leave the halo, while in EAGLE SN-driven outflows travel out to several times the virial radius and act as ejective feedback, draining the halo of its baryons (only a fraction of the EAGLE outflows remain in the halo, where they can cool again and fall onto the galaxy; P. D. Mitchell et al. 2020b). Comparing AGN-dominated halos at the ISM scale, we find that in TNG100 the kinetic AGN feedback is primarily ejective rather than preventative, while the thermal AGN feedback implemented in EAGLE acts in a more preventative manner, suppressing star formation without driving strong ISM outflows.

The mass loading factors of outflows in EAGLE galaxies have a similar mass dependence to those of TNG100, but are lower by a factor of ∼10. While in EAGLE the onset of AGN feedback leads to a flattening of $\eta_M$ at roughly ∼1 for $M_{vir} > 10^{12} M_\odot$, in TNG the mass loading factor increases rapidly as kinetic AGN feedback kicks in—aligning with the dip in both baryon fractions and mass flow rates near this characteristic mass (P. D. Mitchell et al. 2020b).

At the halo scale, the difference between the two simulations is more noticeable. In TNG100, the onset of kinetic AGN feedback acts as both preventative and ejective, delaying inflows from entering the halo in addition to ejecting halo baryons, whereas for higher masses we find the kinetic AGN feedback to be mainly preventative. In EAGLE, P. D. Mitchell et al. (2020b) report only preventative feedback at the halo scale. For a detailed side-by-side comparison of inflow and outflow rates on galaxy and ISM scales in TNG100, EAGLE, and SIMBA, as well as a comparison of the baryon fractions, please see R. J. Wright et al. (2024).

An analysis of how AGN feedback impacts galaxy evolution has also been made using the Horizon simulation (Y. Dubois et al. 2014) by R. S. Beckmann et al. (2017). The AGN feedback model in the Horizon-AGN simulation is similar to that of TNG100, with kinetic and thermal AGN feedback modes separated by the BH accretion rate, such that an AGN starts in the thermal mode and transitions to the kinetic mode as the BH becomes more massive. In the Horizon-AGN simulation R. S. Beckmann et al. (2017) also find that the onset of kinetic AGN feedback quenches galaxies, drives large-scale outflows that significantly reduce the mass of the CGM, and acts as a source of preventative feedback that disrupts gas inflows into galaxies.

### *8.4. Caveats in This Analysis*

#### *8.4.1. Impact of Subgrid Implementations*

The goal of this work has been simply to characterize the flow rates of mass, energy, and metals on galaxy and halo scales in the widely used IllustrisTNG simulation. However, as already touched upon, these quantities can be extremely sensitive to the manner in which the subgrid processes of SN-driven winds and AGN feedback are implemented. This is the reason for the fairly large discrepancies between many aspects of the baryon cycles seen in zoom-in and large volume cosmological simulations, as discussed in more detail in V. Pandya et al. (2020) and R. J. Wright et al. (2024).

There are ongoing efforts to improve these subgrid implementations using insights from smaller-scale, higher-resolution, idealized simulations that more directly resolve the launching of large-scale outflows by SN explosions (e.g., S. Walch et al. 2015; D. Fielding et al. 2017a, 2017b, 2018; C.-Y. Hu 2019; C.-G. Kim et al. 2020a, 2020b; M. Li & G. L. Bryan 2020; E. E. Schneider et al. 2020). An important insight from these simulations is that emergent outflows are multiphase (like observed outflows in real galaxies), with the hot diffuse phase carrying most of the energy (and metals) and the cold dense phase carrying most of the mass. In most past cosmological simulations such as IllustrisTNG and EAGLE, outflows at launch have a single velocity and/or temperature, and typically inject either (mostly) kinetic energy (TNG; R. Weinberger et al. 2017, A. Pillepich et al. 2018a) or only thermal energy (EAGLE; J. Schaye et al. 2015). Thus the predicted baryon cycle emerging from these simulations is almost certainly incomplete.

There are ongoing efforts to develop and implement new multiphase subgrid wind recipes into cosmological simulations, such as the Arkenstone project (J. S. Bennett et al. 2025; M. C. Smith et al. 2024a, 2024b). J. S. Bennett et al. (2025) pointed out that the required mass loadings at launch adopted by A. Pillepich et al. (2018a) in order to match the stellar mass to halo mass relation and other observables in the TNG wind implementation are much higher than those required in the Arkenstone model, when it is implemented within the AREPO code with all other physical processes identical to the TNG physics model. This is because Arkenstone's more accurate treatment of the wind energy coupling with the CGM enables more effective regulation of CGM cooling via preventative feedback, just as predicted by the gas regulator and SAMs of C. Carr et al. (2023), G. M. Voit et al. (2024b), and V. Pandya et al. (2023).

Similarly, the scales on which AGN feedback operates and the manner in which it regulates star formation and BH growth is quite sensitive to the way that BH seeding, accretion, and feedback are implemented (again, see R. J. Wright et al. 2024, for a detailed discussion of the baryon cycle in AGN-dominated halos in TNG, SIMBA, and EAGLE). For example, TNG, like most cosmological simulations, only seeds BHs in halos above a critical mass. However, there is mounting evidence for the existence of both BH- and AGN-driven outflows in lower-mass galaxies (C. M. Manzano-King et al. 2019). A second example is the criterion for switching on the powerful kinetic AGN feedback mode, which depends on both accretion rate and BH mass. This critical BH mass for the onset of kinetic BH feedback creates a "dip" in the CGM mass fraction at intermediate halo masses, which has been correlated with a minimum in the integrated thermal Sunyaev–Zel'dovich signal from the CGMs of galaxies, as shown by Y. Oren et al. 2024), and X-ray brightness of the CGM surrounding galaxies of different masses (see, e.g., E. T. Lau et al. 2025). This feature is not seen in other simulations that implement AGN feedback differently (e.g., EAGLE).

#### 8.4.2. Determining the Dominant Feedback Mechanism

Our analysis of the dominant feedback mechanism in Section 3.3 is key to our interpretation of flow rates in and out of TNG100 galaxies and halos. There are two simplifying assumptions we made in our analysis which may affect our results.

First, in our estimation of the timescales between which we integrate the instantaneous energy injection rates, we assumed outflows do not decelerate as they travel from their source to the halo shell. This work itself has suggested this assumption is incorrect by concluding that as redshifts increase, outflows are more likely to stall as they interact with the dense CGM. Second, as we estimate the total energy injected by each feedback mechanism, we integrate over the instantaneous energy injection rates estimated per snapshot, and over timescales determined by the distance between them. In practice, the time steps between which TNG100 particles evolve are much smaller than the time difference between snapshots. Additionally, kinetic AGN feedback is not injected per time step like thermal AGN feedback and SN feedback, but is accumulated until reaching a certain threshold and is then released. In our analysis we treated all three feedback mechanisms as if they are all injected in all available time steps, which may lead to an inaccurate estimate of when kinetic AGN feedback was injected.

Despite that, we opted to calculate the integrated energies as we did due to the fact that we only use them to compare between mechanisms, and not to estimate any of our quantitative results. More importantly, we integrate the energy injection rates over three different timescales and take the dominant feedback mechanism to be the one that injected more energy throughout two out of three, decreasing the chance of skewing our results due to mistiming.

## 9. Conclusions

We have analyzed the mass, energy, and metal inflow and outflow rates of 9522 TNG100 galaxies and their surrounding halos spanning 5 orders of magnitude in halo virial mass and from $z = 10$ to $z = 0$. We separate our sample into halos with the dominant feedback mechanism estimated to be either SN feedback, thermal AGN feedback, or kinetic AGN feedback. We evaluated the flow rates through an "ISM shell" ($0.05$–$0.15 R_{\rm vir}$) and a "halo shell" ($0.95$–$1.05 R_{\rm vir}$). In addition to measuring the direction of the net flows, we identified markers of preventative versus ejective feedback, and measured the metal enrichment of the flowing gas relative to its source, and the loading factors associated with each type of flow. Our conclusions are as follows.

- (i) The dominant feedback mechanism transitions from SN to kinetic AGN feedback for galaxies in the centers of halos with $M_{\rm vir} \approx 10^{12}\, M_\odot$ at $z = 2$. While thermal AGN feedback, which kicks in at $M_{\rm vir} \approx 10^{11}\, M_\odot$, injects more energy into galaxies compared to SN feedback, it does not appear to significantly affect star formation or drive any outflows.
- (ii) Mass outflow rates (normalized by the halo virial mass) in SN-feedback-dominated halos decrease by 1–1.5 dex from $z \sim 10$ to 0 on halo scales, and by $\sim$2 dex on ISM scales. Normalized energy flow rates show a very similar pattern.
- (iii) Halos of all masses show strongly net positive mass inflow rates at high redshift ($z \gtrsim 2$) on halo scales; on ISM scales the inflow remains net positive in halos with $M_{\rm vir} \lesssim 10^{12}\, M_\odot$, but the excess is much smaller.
- (iv) At high redshift ($z \gtrsim 2$), there is little evidence for halo-scale preventative feedback, with the halo accretion rates modestly exceeding the DM halo growth rate times the universal baryon fraction. Halo-scale preventative feedback remains modest at lower redshift.

(v) On ISM scales, there is evidence for rapid recycling of ejected gas ("galactic fountains") in intermediate- to low-mass halos ($M_{\rm vir} \lesssim 10^{12} M_\odot$).

(vi) The mass loading factor in SN-dominated halos at $z = 0$ decreases from $\eta_M \simeq 20$–5.6 over the mass range $10^{11} \lesssim M_{\rm vir} \lesssim 10^{12.5} M_\odot$. The mass loading at high redshift ($z \sim 2$–10) has a lower normalization and a steeper decline with increasing mass. In contrast, the energy loadings remain nearly constant with both halo mass and redshift. Metal loadings are nearly independent of halo mass, but increase with decreasing redshift.

(vii) At halo masses of $\sim 10^{14} M_\odot$ and above, the gravitational potential begins to become strong enough to resist even the powerful kinetic-mode AGN feedback, leading to a strong increase in the CGM and baryon fractions toward higher halo masses.

(viii) The emergent mass outflow rates do not trace the energy injection rates adopted in the TNG wind subgrid model in a straightforward manner. This may be because outflows are more likely to be stalled by high-density CGM at high redshifts, and to mix with and entrain low-density CGM at low redshifts.

(ix) The metallicities of outflows on halo scales are similar to those of the CGM, or slightly underenriched at very high halo masses ($\gtrsim 10^{13.5} M_\odot$). The metallicities of ISM-scale outflows are similar to the assumed TNG metal enrichment at launch ($Z_{\rm out} \sim 0.4 Z_{\rm ISM}$) for halos where SN feedback is dominant, and increases rapidly to unity for AGN-feedback-dominated halos. The metallicity of gas that is flowing into the ISM is enriched relative to the overall average CGM metallicity at all halo masses, consistent with the measured radial metallicity gradients in the CGM of TNG halos.

The analysis presented in this paper can provide greater insight into why stellar and AGN feedback must behave in certain ways as a function of halo mass and cosmic time in order to reproduce basic quantities such as the galaxy stellar mass fraction, CGM baryon fraction, and quenched galaxy fractions. In turn, this will help guide the development of more physically motivated subgrid recipes and next-generation SAMs.

## Acknowledgments

We wish to thank Jonathan Stern and Amit Nestor-Shachar for their helpful advice and discussions. This work was supported by the German Science Foundation via DFG/DIP grant STE/1869-2GE/625 17-1, by the Center for Computational Astrophysics (CCA) of the Flatiron Institute, and by the Mathematical and Physical Sciences (MPS) division of the Simons Foundation, USA. Support for V.P. was provided by NASA through the NASA Hubble Fellowship grant HST-HF2-51489, awarded by the Space Telescope Science Institute, which is operated by the Association of Universities for Research in Astronomy, Inc., for NASA, under contract NAS5-26555. V.P. thanks Drummond Fielding, Daniel Anglés-Alcázar, and Greg Bryan for getting him interested in gas thermodynamics.

## Data Availability

The data presented in this analysis will be made available upon reasonable request.

## ORCID iDs

Yossi Oren https://orcid.org/0009-0001-3658-6950
Viraj Pandya https://orcid.org/0000-0002-2499-9205
Rachel S. Somerville https://orcid.org/0000-0002-6748-6821
Shy Genel https://orcid.org/0000-0002-3185-1540
Osase Omoruyi https://orcid.org/0000-0002-3649-5362
Amiel Sternberg https://orcid.org/0000-0001-5065-9530